\documentclass[12pt,a4paper]{article}
\usepackage{a4wide}
\usepackage{graphicx,amsfonts,amssymb,epsfig,fancybox,lscape}
\begin{document} {\renewcommand{\thefootnote}{\fnsymbol{footnote}}
\hfill FSUJ-TPI-01/07\\
\mbox{}\hfill CGPG--01/12--1\\
\bigskip
\begin{center} {\LARGE Poisson Geometry in Constrained Systems}\\
\vspace{2em} Martin Bojowald$^a$\footnote{e-mail address: {\tt
bojowald@gravity.phys.psu.edu}} and Thomas Strobl$^b$\footnote{e-mail
address: {\tt Thomas.Strobl@tpi.uni-jena.de}}
\\\vspace{1em} $^a$ Center for Gravitational Physics and Geometry,
Department of Physics,\\ The Pennsylvania State University, University
Park, PA 16802, USA
\\\vspace{0.5em} $^b$Institut f\"ur Theoretische Physik, Universit\"at
Jena, D--07743 Jena, Germany\\
\vspace{3em}
\end{center} }

\newcommand{\id}{{\rm 1\:\!\!\! I}}
\newcommand{\R}{{\mathbb R}}
\newcommand{\N}{{\mathbb N}}
\newcommand{\Z}{{\mathbb Z}}
\newcommand{\Q}{{\mathbb Q}}
\newcommand{\C}{{\mathbb C}}
\newcommand{\CO}{{\cal O}}

\def\ba{\begin{eqnarray}}
\def\ea{\end{eqnarray}}
\def\be{\begin{equation}}
\def\ee{\end{equation}}
\def\re{(\ref }

\newcommand{\Kern}{\mathop{\mathrm{ker}}}
\newcommand{\rank}{\mathop{\mathrm{rank}}}
\newcommand{\Ann}{\mathop{\mathrm{Ann}}\nolimits}

\let\a=\alpha \let\b=\beta \let\g=\gamma \let\d=\delta
\let\e=\varepsilon \let\ep=\epsilon \let\z=\zeta \let\h=\eta
\let\th=\theta
\let\dh=\vartheta \let\k=\kappa \let\l=\lambda \let\m=\mu
\let\n=\nu \let\x=\xi \let\p=\pi \let\r=\rho \let\s=\sigma
\let\t=\tau \let\o=\omega \let\c=\chi \let\ps=\psi
\let\ph=\varphi \let\Ph=\phi \let\PH=\Phi \let\Ps=\Psi
\let\O=\Omega \let\S=\Sigma \let\P=\Pi
\let\Th=\Theta \let\L=\Lambda \let \G=\Gamma \let\D=\Delta
\def\wtO{\widetilde{\Omega}}

\def\lb{\left\{} \def\rb{\right\}}
\let\lra=\leftrightarrow \let\LRA=\Leftrightarrow
\def\ul{\underline}
\let\Ra=\Rightarrow \let\ra=\rightarrow
\let\la=\leftarrow \let\La=\Leftarrow

\def\CG{{\cal G}}\def\CN{{\cal N}}\def\CC{{\cal C}}
\def\CL{{\cal L}} \def\CX{{\cal X}} \def\CA{{\cal A}} \def\CE{{\cal
E}}
\def\CF{{\cal F}} \def\CD{{\cal D}} \def\rd{\rm d}
\def\rD{\rm D} \def\CH{{\cal H}} \def\CT{{\cal T}} \def\CM{{\cal M}}
\def\CI{{\cal I}}
\def\CP{{\cal P}} \def\CS{{\cal S}} \def\C{{\cal C}}
\def\CR{{\cal R}}
\def\CO{{\cal O}}
\def\CU{{\cal U}}

\newcommand{\md}{\mathrm{d}}
\newcommand{\mdp}{\mathop{{\mathrm{d}}_{\parallel}}\nolimits}
\newcommand{\mdo}{\mathop{{\mathrm{d}}_{\perp}}\nolimits}
\newcommand{\Diff}{\mbox{\rm Diff}}
\newcommand{\diff}{\mbox{\rm diff}}

\newtheorem{theo}{Theorem}
\newtheorem{lemma}{Lemma}
\newtheorem{cor}{Corollary}
\newtheorem{defi}{Definition}
\newtheorem{prop}{Proposition}

\newcommand{\proofend}{\raisebox{1.3mm}{%
\fbox{\begin{minipage}[b][0cm][b]{0cm}\end{minipage}}}}
\newenvironment{proof}[1][\hspace{-1mm}]{{\noindent\it Proof #1:}
}{\mbox{}\hfill \proofend\\\mbox{}}

\newenvironment{rem}{{\noindent\it Remark:}}{\\\smallskip}
\newenvironment{ex}{{\noindent\it Example:}}{\\\smallskip}

\newcommand{\tom}{\widetilde \omega}
\newcommand{\Tom}{\widetilde \Omega}

\newcommand{\2}{\frac{1}{2}}
\newcommand{\wt}{\widetilde}
\def\wtO{\wt \Omega}

\newcommand{\ie}{i.e.\ }
\newcommand{\cf}{cf.\ }
\newcommand{\ins}{\lrcorner}
\newcommand{\emb}{\hookrightarrow}

\begin{abstract} Associated to a constrained system with closed
constraint algebra there are two Poisson manifolds $P$ and $Q$ forming
a symplectic dual pair with respect to the original, unconstrained
phase space: $P$ is the image of the constraint map (equipped with the
algebra of constraints) and $Q$ the Poisson quotient with respect to
the orbits generated by the constraints (the orbit space is assumed to
be a manifold). We provide sufficient conditions so that the reduced
phase space of the constrained system may be identified with a
symplectic leaf of $Q$. By these methods, a second class constrained
system with closed algebra is reformulated as an {\em abelian\/} first
class system in an extended phase space.

While any Poisson manifold $(P,\Pi)$ has a symplectic realization
(Karasev, Weinstein 87), it does not always permit a leafwise
symplectic embedding into a symplectic manifold $(M,\o)$.  For regular
$P$, it is seen that such an embedding exists, iff the characteristic
form-class of $\Pi$, a certain element of the third relative
cohomology of $P$, vanishes. A tubular neighborhood of the constraint
surface of a general second class constrained system equipped with the
Dirac bracket provides a physical example for such an embedding into
the original symplectic manifold. In contrast, a leafwise symplectic
embedding of e.g.\ (the maximal regular part of) a Poisson Lie
manifold associated to a compact, semisimple Lie algebra does not
exist.

\end{abstract}

\newpage
\setcounter{tocdepth}{3}
\tableofcontents

\setcounter{footnote}{0}

\section{Introduction}

Within constrained Hamiltonian mechanics one is used to the concept of
presymplectic manifolds, manifolds equipped with a closed 2-form.
Pulling back the symplectic form $\omega$ from the original phase
space $M$ to the constraint surface $C$, the resulting 2-form
$\omega_C$ is still closed, but in general no longer
nondegenerate. The constraints (functions on $M$ characterizing the
constraint surface $C$ as joint preimage of zero) are of `second
class' \cite{Dirac}, iff $\omega_C$ happens to be nondegenerate;
$(C,\omega_C)$ is then the reduced phase space of the
theory. Otherwise, there are vector fields in the kernel of
$\omega_C$, which are integrable as a consequence of the Frobenius
theorem; their orbits are the `gauge transformations' of the
theory. Taking the factor space, provided well-defined, yields the
reduced (or physical) phase space in this more general case.

Recently there has been increasing interest in Poisson manifolds, in
part because of its relation to deformation quantization (\cf e.g.\
\cite{Kontsevich,SW,Vaisman}) and the interplay of String Theory with
noncommutative gauge theories (\cf e.g.\ \cite{Schomerus,SchuppWess}).

Poisson manifolds are a generalization of symplectic manifolds in a
way dual to (but different from) the one of presymplectic manifolds:
Instead of defining a symplectic manifold through the existence of a
nondegenerate 2-form $\omega = \2 \omega_{ij} \md x^i \wedge \md x^j$,
closed due to Jacobi, one could as well define it by its (negative)
inverse, i.e.\ by a nondegenerate bivector field $\Pi = \2\Pi^{ij}
\partial_i \wedge \partial_j$, where $\Pi^{ij} = -(\omega^{-1})^{ij}
\equiv \{ x^i,x^j \}$, the fundamental Poisson brackets between local
coordinates.\footnote{The reason for the somewhat conventional minus
sign in front of the matrix inverse to $\o_{ij}$ in the definition of
$\Pi^{ij}$ will be commented on in Footnote \ref{footnote5}.} Now the
Jacobi identity takes the form $[\Pi,\Pi]\equiv -
\Pi^{ij}{}_{,s}\Pi^{sk} \partial_i \wedge \partial_j \wedge \partial_k
= 0$. While presymplectic manifolds result from giving up the
nondegeneracy of $\omega_{ij}$ (keeping $\md \o =0$), Poisson
manifolds result from giving up nondegeneracy of $\Pi^{ij}$ (keeping
$[\Pi,\Pi]=0$).  In the degenerate (nonsymplectic) Poisson case,
Hamiltonian vector fields $v_f := -\md f \ins \Pi \equiv \{\cdot,f \}$
do no longer span the full tangent space $T_xP$ at any $x\in
P$. Still, the respective distribution is integrable (since
$[v_f,v_g]=v_{\{g,f\}}$ due to Jacobi, \ie due to $[\Pi,\Pi]=0$),
generating a foliation (stratification) of $P$ into symplectic leaves.

In this paper we show that also the notion of Poisson manifolds plays
some role within the physical scenario of constrained Hamiltonian
systems. We show how (also nonsymplectic) Poisson manifolds arise
naturally in this context and how they are related to the respective
(pre)symplectic manifolds of the system.

The first instance, where Poisson manifolds show up, is in the case
that the Poisson algebra of the constraint functions defines a closed
Poisson subalgebra.  As pointed out recently \cite{ASS}, such a
constrained system does not only give rise to a presymplectic manifold
(the constraint surface $C$ embedded in the original symplectic
manifold) but is equally naturally associated to a Poisson manifold
$P$, $P$ being the image of the original phase space $M$ under the
constraint map, endowed with the Poisson algebra of these
constraints. In fact, invoking theorems from the mathematics
literature, \emph{any} Poisson manifold\footnote{In the context of
physics, constraints are usually globally defined functions on $M$ and
then only any region of $P$ homeomorphic to a subset of
$\R^d$, $d = \mbox{dim} (P)$, can be obtained. For most of our
considerations, however, we can relax this condition on the
constraints.} $P$ or {\em any\/} presymplectic
manifold $C$ (though not any pair $(P,C)$) can be regarded as arising
in this manner (cf.\ Theorem \ref{realization} and
\ref{Gotay} below).

Since the constraints form a closed algebra, their Hamiltonian vector
fields define an integrable distribution (on {\em all\/} of $M$, not
just on $C$ as one is used to from a general first class constrained
system). Assume the quotient of $M$ with respect  to these
`generalized gauge orbits' gives a well-defined, differentiable
manifold  $Q$. Then also $Q$ can be endowed with a natural Poisson
bracket; essentially it is just the original Poisson algebra of `gauge
invariant' functions on $M$.

In Sec.~\ref{sec:Closed} of this paper we refine this construction. In
particular, we show that $P$ and $Q$ form a so called symplectic dual
pair with respect to the original phase space $M$ 
and specify conditions (\cf Theorem \ref{dual}, Corollary 2', and
Theorem \ref{main} below) such that the reduced phase space $R$, of
primary physical interest, may be identified with a symplectic leaf of
$Q$.
                                    
For a closed first class constrained system (\ie the origin in $P$ is a
symplectic leaf) this implies that we may reverse the two steps in the
reduction process: Instead of first restricting to a submanifold
(which is presymplectic) and then taking the factor space with respect
to the gauge orbits, we may as well first take a factor space with
respect to generalized gauge orbits, leading to the Poisson manifold
$Q$, and only then take the restriction (in this case to a particular
symplectic leaf).

Moreover, by these means also a closed second class constrained system
may be viewed similarly to a (general) first class constrained
system. This will be made more explicit in Sec.~\ref{subsecAppl} by
embedding the original, unconstrained phase space $M$ into some
extended phase space $\widetilde M$, where for any second class
constraint function in $M$ one gets a first class constraint function
in $\widetilde M$. We will show that upon an appropriate extension,
one can manage to have the first class constraints to Poisson
commute. This may constitute a significant simplification for the
quantization of the original system.

There is also another instance where Poisson manifolds are used in the
context of constrained systems. In the case of second class
constraints $\Phi^\alpha \approx 0$ (not necessarily closed under
taking Poisson brackets), it was Dirac himself who introduced a
modified Poisson bracket on the original phase space, the Dirac
bracket. This bracket has the feature that the constraints
$\Phi^\alpha(x)$ are its Casimir functions\footnote{A function $C$ is
called a Casimir function of a Poisson tensor $\Pi$ if
$\Pi^{ij}\partial_j C=0$.} and, essentially, that restricting to the
constraint surface commutes with taking Poisson brackets. In other
words, one defines a new Poisson bivector $\Pi_D$ such that one of its
symplectic leaves coincides (as a symplectic manifold) with the
reduced phase space $\Phi^\alpha = 0$ of the theory. In fact, such a
relation holds even for a whole neighborhood of leaves: The slightly
deformed constraint surfaces $\Phi^\alpha = c^\alpha$, for constants
$c^\alpha$ out of some interval containing zero, endowed with the
symplectic form inherited from the embedding original phase space
$(M,\o)$, are symplectic leaves of $\Pi_D$.

In general $\Pi_D$ is defined only in some tubular neighborhood $S$ of
the constraint surface $C$. What is the relation between $(M,\o)$ and
$(S,\Pi_D)$? Clearly, the restriction of the identity map on $M$ to
$S$ is {\em not\/} a Poisson map since the respective Poisson brackets
coincide only for particular functions.  On the other hand, the
embedding map from $S$ to $M$ is leafwise symplectic
(or {\em leaf-symplectic}), \ie restricted to any leaf of $\Pi_D$, the
map is symplectic.  

In Sec.~\ref{sec:LeafSymp} we consider the question whether a given
Poisson manifold $(P,\Pi_P)$ admits a leaf-symplectic embedding into
some symplectic manifold.  Locally and around a regular point $x$ of
$P$ such a map exists always: In a neighborhood $N$ of $x$ one may
choose Casimir--Darboux coordinates $(q^\alpha,p_\beta,C^I)$, $I = 1,
\ldots , k$ where $k$ is the corank of $\Pi$ at $x$, for which the
given bivector has the form $\Pi_P = \frac{\partial}{\partial
q^\alpha} \wedge \frac{\partial}{\partial p_\alpha}$ \cite{We83}. The
respective embedding phase space may be chosen as $(N \times \R^k, \md
p_\alpha \wedge \md q^\alpha + \md P_I \wedge \md C^I)$, where $P_I$
are linear coordinates in $\R^k$ and the embedding corresponds to
fixing these `Casimir momenta' to some value.

Globally, however, there can be obstructions which may be
characterized by a closed 3-form on $P$, the so called form-class
\cite{VaismanBook} of the bivector $\Pi_P$. Considered as an element
of the third relative cohomology of a regular Poisson manifold
$(P,\Pi_P)$, its vanishing is necessary and sufficient for the
existence of a leaf-symplectic embedding (\cf Proposition
\ref{prop:ls} below). The condition on the form-class may,
furthermore, be cast into the form of descent equations, familiar to
physicists from the analysis of anomalies (cf, e.g., \cite{Bertlmann})
and, more recently, also from the cohomological deformation of action
functionals \cite{Barnich}.  In this manner we will find, e.g., that a
family of coadjoint orbits of a compact, semisimple Lie algebra
(viewed as Lie Poisson manifold) does not permit a leaf-symplectic
embedding. On the other hand, any regularly foliated Poisson manifold
whose leaves have trivial second cohomology does so.

Thus while there always exists a surjective Poisson map from a
symplectic manifold to any given Poisson manifold
\cite{SympReal,Weinstein} (i.e., a so called symplectic realization,
\cf Theorem 1 below),
and likewise always a coisotropic embedding of (regular) presymplectic
manifolds into a symplectic manifold \cite{Gotay,GS} (Theorem 2
below), a leaf-symplectic embedding of a given Poisson manifold
exists only in particular cases.

Both relations of (pre)symplectic and Poisson manifolds studied in
this paper have physical applications: The information in the
symplectic dual pairs, and in particular the reformulation of a second
class constrained system with closed algebra as an abelian first class
system, can be used, e.g., in a path integral quantization, the prime
example possibly being some Yang-Mills gauge theory where the
constraint algebra is spoiled by an anomaly (\cf also
\cite{FS,Batalin,Batalin2} for related work). The existence of
leaf-symplectic maps for a regular Poisson manifold leads to solutions
\cite{PSMclass,Habil} of the associated Poisson Sigma model
\cite{PSM,Ikeda} on a Riemann surface.

Before describing the new results we set the stage
in Sec.~2, recalling the
basic definitions of different types of submanifolds in a
symplectic manifold, in particular the definitions of
first and second class
constraint surfaces going back to Dirac. On this occasion we also
suggest generalizations of these notions to submanifolds of
presymplectic and Poisson manifolds (Def.~\ref{coisoPoi}).

\section{Preliminaries---Different types of constraints} 
\label{Prelim}

We start with some general remarks on constrained systems.  Let us
consider a phase space, i.e.\ a symplectic manifold, $(M,\omega)$ in
which we single out a submanifold $C$, called the constraint surface.
We assume that this submanifold can be characterized as the
intersection of the zero level set of $\{\Phi^\alpha \in C^\infty(M),
\: \alpha = 1, \ldots , d \}$, the constraint functions or simply the
constraints. We further assume that the constraints are regular and
irreducible, which means that $\bigwedge_{\alpha = 1}^d \md
\Phi^\alpha$ does not vanish on $C$. The Hamiltonian of the system
Poisson commutes with all the $\Phi^\alpha$, \ie the set of
$\Phi^\alpha$s is already the total set of constraints---we are not
interested in a splitting into ``primary'' and ``secondary''
constraints \cite{Dirac} as it arises when starting from different
Lagrangian systems, eventually leading to the same constrained
Hamiltonian system.

We will, however, use the notion of first class and second class
constraints:
\begin{defi} \label{constr}
 A constraint function $\Phi^\alpha$ of a constrained system $C
 \hookrightarrow (M,\omega)$ is of the {\em first (second) class\/},
 if its Hamiltonian vector field $v^\alpha \equiv \{ \cdot,
 \Phi^\alpha \}$ is (is nowhere) tangent to $C$. The full set of
 constraint functions $\{ \Phi^\alpha \}$, such that $C =
 \bigcap_\alpha (\Phi^\alpha)^{-1}(0)$, is of the first (second)
 class, if each single constraint function $\Phi^\alpha$ is of the
 first (second) class.
\label{defi1}
\end{defi} 
In other words, $\Phi^\alpha$ is of first (second) class, iff
$v^\alpha$ lies (does not lie) in $\iota_*TC$ for {\em all\/}
points on the constraint surface $C \subset M$,
$\iota\colon C\to M$ denoting the respective embedding map. 

In general a constraint will be
neither of first nor second class, but of `mixed type', and a
splitting of constraints, characterizing a given constraint surface $C
\subset M$, into first and second class constraints can be achieved
only locally.

The definition as given above is readily seen to reproduce the one
given by Dirac \cite{Dirac} (use $\{ \Phi^\alpha , \Phi^\beta \}|_C
\equiv v^\beta (\Phi^\alpha)$ or \cf Proposition 1 below): A system of
constraints $\{ \Phi^\alpha \}$ is of the first class, iff $\{
\Phi^\alpha , \Phi^\beta \}|_C =0$ $\forall \alpha, \beta$, and of the
second class, iff $\det \{ \Phi^\alpha , \Phi^\beta \}|_C \neq 0$.

In the more mathematically inclined literature, the above two types of
constraint surfaces are characterized in a different way, without any
reference to the constraint functions. First note that the pullback
bundle $\iota^* TM$ (or $TM|_C$) is a symplectic vector bundle
\cite{Vaismansymp}\footnote{\label{footnote3}
Let us call a vector bundle endowed with a
differentiable field of bilinear antisymmetric forms of its fibers a
{\em presymplectic vector bundle}.  If this bilinear form is
nondegenerate, the bundle is symplectic \cite{Vaismansymp}.  Note that
both the tangent bundle $TM$ of a presymplectic manifold $(M,\o)$ and
the cotangent bundle $T^*M$ of a Poisson manifold $(M,\Pi)$ are
presymplectic vector bundles over $M$ in this terminology. In this
context it is worthwhile mentioning a possible unification of the two
apparently different generalizations of symplectic manifolds 
in terms of so called
{\em Dirac structures\/} \cite{Diracstructures1,Diracstructures2}.
Here one starts from $TM \oplus T^*M$, endowed with the canonical
symmetric bilinear form from the pairing and looks for certain  
maximally isotropic subbundles. (By definition a subbundle
in a vector bundle equipped with a nondegenerate symmetric
or antisymmetric bilinear form $\kappa$ on its fibers
is isotropic if all of its sections are mutually orthogonal with
respect to $\kappa$). Although it may be
worthwhile to explore the considerations of the present paper within
this more general framework of Dirac structures, we will not attempt
to do so here.}
  over C and $\iota_* TC$ (or simply $TC$) a subbundle
thereof. This subbundle induces a symplectically orthogonal subbundle
\be (TC)^{\perp}:=\{v\in TM|_C:\o(v,w)=0 \mbox{ for all }w\in TC\}
\label{TCnormal} \ee which is related to the annihilator of $TC$ in
$M$ (or the ``conormal bundle'' of $C$), \be
\mathrm{Ann}_M(TC)=\{\alpha\in T^*M|_C:\alpha(v)=0 \mbox{ for all
}v\in TC\} \, , \label{Ann} \ee in the following way. Denoting the map
from a vector space $V$ to its dual $V^*$ induced by a bilinear form
$B$ on $V$ as $B^\sharp$, $v \mapsto B(v,\cdot)$, the symplectic form
$\omega \in \Omega^2(M)$ defines a bijection $\omega^\sharp \colon TM
\to T^*M$. Denote by $\Pi^\sharp \colon T^*M \to TM$ the transpose
inverse map, $\Pi^\sharp :=(\omega^\sharp)^{-1,T} \equiv
-(\omega^\sharp)^{-1}$, which implicitly defines a bivector field
$\Pi$.\footnote{\label{footnote5} Given an arbitrary, fiberwise
nondegenerate bilinear form $\omega$ on $TM$, $\omega^\sharp$ defines
a bijection which may be lifted to the full tensor bundle over $M$ and
which thus allows to identify different types (covariant,
contravariant) of tensors (``lowering and raising of indices'').  If
we want $\Pi \in \Gamma(TM \otimes TM)$ to {\em be the same\/} object
as $\omega \in \Gamma(T^*M \otimes T^*M)$ in this identification,
$\Pi=\left((\omega^\sharp)^{-1}\otimes (\omega^\sharp)^{-1}\right)
(\omega)$, we find $\Pi^\sharp =
(\omega^\sharp)^{-1,T}$. (Pseudo-)Riemannian geometry, where the
bilinear form $\omega \equiv g$ is symmetric, is special in two ways:
The transpose is irrelevant, $g^{ij}$ is the inverse to $g_{ij}$, and,
maybe more important, symmetry of the bilinear form is the only case
where index transport commutes with contraction of tensors. In the
antisymmetric case considered here, we thus find the additional minus
sign in the definition of $\Pi^{ij}$; moreover, now e.g.\ $A_i B^i = -
A^i B_i$. (See also Appendix B of \cite{Urbantke} for a discussion).}
We now find $\Pi^\sharp \left(\mathrm{Ann}_M(TC)\right)=(TC)^{\perp}$
or $\mathrm{Ann}_M(TC)= \omega^\sharp \left((TC)^{\perp}\right)$: The
inclusion is established by noting that $\alpha\in\Ann_M(TC)$ implies
$\omega(\Pi^\sharp\alpha,w)=-\alpha(w)=0$ for any $w\in TC$, while,
vice versa, $\omega(v,w)\equiv \omega^\sharp v(w)=0$ for all $w\in TC$
implies $\omega^\sharp(TC)^{\perp} \subset \mathrm{Ann}_M(TC)$.

Clearly, $\mathrm{Ann}_M(TC) = \mbox{$\prec\!\!\md \Phi^\alpha\!\!
\succ$}\,$ since $\,\md \Phi^\alpha (w) = w(\Phi^\alpha)$ vanishes for
any tangent vector $w \in TC$. With $v^\alpha = - \md \Phi^\alpha \ins
\Pi \equiv - \Pi^\sharp \md \Phi^\alpha$ we similarly recognize
$(TC)^{\perp}$ as the span of the Hamiltonian vector fields generated
by the constraints: $(TC)^{\perp} = \,\mbox{$\prec\!\!v^\alpha \!\!
\succ$}$.

We now have several equivalent characterizations of first and second
class constrained systems:

\begin{prop} \label{FirstSecondSymp}
 For a constrained system $C \hookrightarrow (M,\omega)$ the
 statements in part (i) and (ii), respectively, are equivalent to one 
 another:
\begin{enumerate}
 \item[(i)]
\begin{enumerate}
 \item $C$ is first class.
 \item $(TC)^{\perp} \subset TC$.
 \item $\Pi|_{\Ann_M(TC)}=0$, \ie
 $\Pi(\alpha,\beta)=0$ for all
 $\alpha,\beta\in \Ann_M(TC)$.
 \item The embedding $\iota\colon C\to M$ is
  coisotropic, \ie $(TC)^{\perp}$ is isotropic $(\o|_{(TC)^\perp}
  =0)$.
\end{enumerate}
\item[(ii)]
\begin{enumerate}
 \item $C$ is second class.
 \item $(TC)^{\perp} \cap TC = 0$.
 \item $\Pi|_{\Ann_M(TC)}$ is nondegenerate.
 \item $\o|_{(TC)^\perp}$ is nondegenerate. 
 \item $C$ is symplectic, \ie $\iota^* \omega$  (or
 $\o|_{TC}$) is nondegenerate.
\end{enumerate}
\end{enumerate}
\end{prop} 

In (ii), (b) $0$ denotes the zero section.

\begin{proof}
Equivalence of (a) and (b) in (i) and (ii) follows immediately from
Definition \ref{defi1} with the preceding considerations. Likewise for
(c) and (d), if we note that $\omega(v^\alpha,v^\beta) = \Pi( \md
\Phi^\alpha,\md \Phi^\beta)$; since, moreover, the righthand side of this
equation is equal to $\{\Phi^\alpha, \Phi^\beta \}|_C$, (c) in (i) and
(ii) is recognized as the original definition of Dirac, formulated,
however, without the use of constraint functions $\Phi^\alpha$ used to
specify the embedded constraint surface $C$.

Concerning (i) it is now sufficient to establish equivalence between 
(b) and (d): $(TC)^{\perp} \subset
TC$ by definition implies isotropy of $(TC)^{\perp}$, and isotropy of
$(TC)^{\perp}$ yields
\[
 \Ann_M(TC)((TC)^{\perp})=
 [\omega^{\sharp}((TC)^{\perp})]((TC)^{\perp})= 0 \; .
\]

Concerning (ii) we next show (b) $ \LRA $ (e): Let us
assume that $\iota^* \omega$ is degenerate, which implies that there
is a point $p\in C$ which has a nonzero tangent vector $v\in T_pC$
such that $\iota^* \omega(v,w)=0$ for all $w\in T_pC$. This is in
contradiction to $TC\cap(TC)^{\perp} = 0$. Likewise,
if there is an isotropic tangent vector, i.e.\ a nonzero vector
which is contained in both $TC$ and $(TC)^{\perp}$, $\iota^* \omega$
is obviously degenerate.

Finally, equivalence between (b) and (d) in (ii) may be shown likewise
by noting that $TC = ((TC)^\perp)^\perp$, so that we may replace $TC$
by $(TC)^\perp$ in the preceding argument. 
\end{proof}

Since we are interested in both presymplectic and Poisson manifolds as
two different generalizations of symplectic manifolds, let us in the
following suggest a generalization of the notions of first class
and second class submanifolds to these cases, ensuring agreement
on their symplectic intersection. In this context we want to keep the
following characteristics of first and second class constraint
surfaces $C$: It should not be possible that $C$ is
simultaneously first and second class (Dirac's classification is not
exhaustive, but at least it should remain exclusive).
Moreover, a second class constraint surface
is a (nondegenerate) phase space of its own ($C$ carries a natural
symplectic structure), and a first class
constraint surface is a phase space after factoring out gauge
transformations which are generated by the flow of the constraints (in
both cases, first and second class,
this defines the reduced phase space).\footnote{For presymplectic
manifolds the flow generated by a constraint is not uniquely
defined. However, the ambiguity is the kernel of $\o \in
\Omega^2(M)$, which we thus want to include into the generators of the
`flow of gauge transformations'. This is achieved in the definition
below.}  On the other hand,
at least for Poisson manifolds there already exists a reasonable
notion of coisotropic submanifolds (cf., e.g., \cite{SW}). In this
way we arrive at the following definition:

\begin{defi}\label{coisoPoi}
\noindent 
\begin{enumerate}
 \item[(i)] Let $C$ be a (closed) submanifold of a presymplectic manifold
 $(M,\omega)$.
\begin{enumerate}
 \item $C$ is called {\em coisotropic}, if
 $\,(TC)^{\perp}$---as defined in (\ref{TCnormal})---is isotropic,
 \item it is called {\em first class} if $\,(TC)^{\perp} \subset TC$, and
 \item it is called {\em second class}, if $\,TC \cap (TC)^{\perp} =
 0$.
\end{enumerate}
 \item[(ii)] A (closed) submanifold $C$ of a Poisson manifold
 $(M,\Pi)$ is called {\em coisotropic}, if $\Ann_M(TC)$ is
 isotropic, \ie $\Pi|_{\Ann_M(TC)}=0$.
 \item[(iii)] Let $C$ be a (closed) submanifold of a Poisson manifold
 $(M,\Pi)$ such that $\iota_* TC \subset \Pi^{\sharp}(T^*M)$.
\begin{enumerate}
 \item $C$ is called {\em first class\/} if
 $\,0\not=\Pi^{\sharp}\Ann_M(TC)|_x\subset T_xC$ for any point $x\in C$,
   and
 \item {\em second class\/} if $\,\Pi^{\sharp}\Ann_M(TC)\cap TC=0$.
\end{enumerate}
\end{enumerate}
\end{defi}

Using the fact that in the symplectic case $\Pi^{\sharp}$ defines an
isomorphism between $\Ann_M(TC)$ and $(TC)^{\perp}$, it is easy to see
that the respective definitions for presymplectic and Poisson
manifolds coincide if $M$ is symplectic. In particular, the additional
condition in part (iii) is always true in this case due to
$\Pi^{\sharp}(T^*M)=TM$.

\medskip

\begin{rem}
 Conditions (a) and (b) in part (i) are not equivalent to one another
 if $(M,\omega)$ is not symplectic. In particular, there may be
 coisotropic second class submanifolds (iff $C$ is a symplectic
 submanifold of $M$ with $(TC)^\perp = \ker \omega$).  On the other
 hand, in the Poisson case $\Pi|_{\Ann_M(TC)}=0$ {\em is\/} equivalent
 to $\Pi^{\sharp}\Ann_M(TC) \subset TC$. Nonetheless, still a
 coisotropic submanifold of a Poisson manifold is not necessarily
 first class because the additional condition in part (iii) may be
 violated. In both cases, Poisson and presymplectic, a first class
 submanifold $C$ is coisotropic, but in general not vice versa.
\end{rem}

It follows as in previous considerations that a first class
submanifold of a presymplectic manifold $(M,\o)$, with the pulled back
2-form $\o_C = \iota^* \o$, is degenerate, i.e.\ the kernel has to be
factored out to arrive at a reduced phase space, whereas a second
class submanifold is always symplectic.

If we have a submanifold $C$ of a Poisson manifold $(M,\Pi_M)$, on the
other hand, it is in general not possible to even {\em define\/} a
Poisson or presymplectic structure on $C$ directly since we cannot use
the pull back as in the (pre)symplectic case. However, one of the
instances where this is possible occurs, if the annihilator is a
Poisson ideal: \be \Pi_M^{\sharp}(\Ann_M(TC))=0,
\label{ideal} \ee i.e.\ $\Pi_M(\alpha,\cdot)=0$ for all
$\alpha\in\Ann_M(TC)$. Using the isomorphism
\[
 i\colon T^*C\to T^*M|_C/\Ann_M(TC)\;,\:\alpha\mapsto [\alpha_M] =
 \alpha_M+\Ann_M(TC)
\]
where $\alpha_M$ is some element of $T^*M|_C$ fulfilling
$\alpha_M(v)=\alpha(v)$ for all $v\in TC\subset TM$,
\begin{equation}\label{PoissStruc}
 \Pi_C(\alpha,\beta):=\Pi_M([\alpha_M],[\beta_M])|_C
\end{equation}
leads to a well-defined Poisson tensor $\Pi_C$ on $C$, if
(\ref{ideal}) is satisfied.

Note that the condition (\ref{ideal}) is
{\em stronger\/} than the one required for coisotropic submanifolds,
which only requires $\Pi^{\sharp}\Ann_M(TC)\subset TC$;
correspondingly a general coisotropic submanifold of a Poisson
manifold does not carry a canonical Poisson (or presymplectic)
structure. The (coisotropic) submanifolds $C$ of smallest possible
dimension fulfilling (\ref{ideal}) are the symplectic leaves; in this
case $\Ann_M(TC) = \ker \Pi^\sharp$ and $\Pi_C$ becomes
nondegenerate. The symplectic form $\o_C$ on the leaf $C$ is then
defined by $(\o_C)^\sharp = ((\Pi_C)^\sharp)^{-1}$. 

The additional condition $\iota_* TC \subset \Pi^{\sharp}(T^*M)$ in
part (iii) of Definition~\ref{coisoPoi} ensures that $C$ is a subset
of a symplectic leaf in $(M,\Pi)$. It is then always possible to
define a symplectic structure on a first class or second class
submanifold $C$ by pulling back the symplectic structure on the leaf
to $C$. In this way, a first class submanifold acquires a degenerate
presymplectic structure, whereas a second class submanifold always
is symplectic.

In the following section we focus on {\em closed\/} constraint
algebras, \ie on a constrained system $(M,\omega, \{ \Phi^\alpha \})$
in which the constraints $\Phi^\a=0$ generate a Poisson subalgebra.
This implies that 
\be \{ \Phi^\a , \Phi^\b \}_M (x) = \Pi_P^{\a\b}(\Phi(x))
\,  \label{consalg} \ee
holds for some $\Pi_P^{\a\b}(\Phi)$. 
In the context of (\ref{consalg}) the constraint
surface $C$ is of first class, iff
$\Pi_P(0) = 0$ and of second class, iff $\det \Pi (0) \neq 0$.

For simplicity we will take $M$ to be connected in the following 
and require that the constraint functions $\Phi^\a$ can be used as
part of a local coordinate system around any point in $M$ (instead of
just points in $C \subset M$, which follows from the regularity and
irreducibility of the constraints required in the beginning of this
section).

\section{Poisson geometry from closed constraint algebras}
\label{sec:Closed} 

In this section we present results on the geometry of
closed constraint algebras and the relation to the reduced phase
space. This then yields an alternative procedure for a constraint
reduction.

\subsection{Symplectic dual pairs and related morphisms}

Define $P$ as the subset of $\R^d$ with coordinates $\Phi^\a$ which is
in the image of $M$ under the constraint map $\phi\colon M\to
P\subset\R^d,x\mapsto\Phi^{\alpha}(x)$ and endow $P$ with the bivector
$\Pi_P=\2 \Pi_P^{\a\b}(\Phi) \partial_\a \wedge \partial_\b$ appearing
in the closed constraint algebra (\ref{consalg}).  In this way we have
\begin{equation} (C,\omega_C)\stackrel{\iota}{\hookrightarrow}
(M,\omega)\stackrel{\phi}{\longrightarrow} (P,\Pi_P) \; .
\label{P} \end{equation}
Both arrows in this diagram are morphisms. However, there are two
different categories involved: $C$ endowed with $\omega_C = \iota^*
\o$ is a presymplectic manifold (as before $\iota$ denotes the
embedding of the constraint surface $C$ into $M$); note that in the
present case the dimension of the kernel of $\omega_C$ is necessarily
constant on $C$, and in the following we will include this condition
on the kernel of the 2-form in the definition of a presymplectic
manifold unless stated otherwise. $(P,\Pi_P)$, on the other hand, is
Poisson. Correspondingly, $\iota\colon C\to M$ is a morphism of
presymplectic manifolds and $\phi\colon M\to P$ a morphism of Poisson
manifolds, where, in the first instance, $(M,\o)$ is regarded as a
(nondegenerate) presymplectic manifold and, in the second case, as a
(nondegenerate) Poisson manifold.
\begin{defi} \label{morphium} A morphism $f$ between two
(pre)symplectic manifolds $(M_1,\o_1)$ and  $(M_2,\o_2)$, a
{\em (pre)symplectic map}, is a map 
$f\colon M_1 \to M_2$ such that $f^* \o_2 = \o_1$.
A morphism $g$ between two
Poisson manifolds $(M_1,\Pi_1)$ and  $(M_2,\Pi_2)$, a {\em Poisson
map}, is a map 
$g\colon M_1 \to M_2$ such that $g_* \Pi_1 = \Pi_2$.
\end{defi}

An alternative characterization of a Poisson map makes use of the
definition of coisotropy (Definition~\ref{coisoPoi}) and the notion of
the graph $\Gamma_g=\{(m_1,g(m_1)):m_1\in M_1\}\subset M_1\times M_2$
of a map $g\colon M_1\to M_2$:

\begin{lemma}[\cite{Graph,SW}] \label{PoissonGraph}
 A map $g\colon M_1\to M_2$ between Poisson manifolds is a Poisson map
 if and only if its graph $\Gamma_g$ is coisotropic in
 $M_1\times\bar{M}_2$ where $\bar{M}_2$ has the negative Poisson
 structure of $M_2$.
\end{lemma}

Equivalently a Poisson map $g$ may be characterized by $g^*
\{F,G\}_{M_2} = \{ g^* F,g^* G \}_{M_1}$ for all $F,G \in
C^\infty(M_2)$.\footnote{Warning: In general (and in particular always
for (\ref{P}) in the instance of symplectic $C$ and $P$, as we will
verify explicitly in subsection \ref{sec:Second} below) a Poisson map
(a (pre)symplectic map) between two symplectic manifolds is {\em
not\/} symplectic (Poisson).}  Thus by construction $\phi\colon M\to
P$ is Poisson (cf.\ e.g.\ (\ref{consalg})) and found to provide a
surjective, submersive symplectic realization of $P$:

\begin{defi} 
 A {\em symplectic realization\/} of a Poisson manifold $P$ is a
 Poisson map $\phi\colon M\to P$ where $M$ is a symplectic manifold.
\end{defi}

For our mathematical considerations we will drop the condition that
$P\subset\R^d$ with $d=\dim P$. We then have to single out a point
$0\in P$ in order to define the constraint surface
$C=\phi^{-1}(0)$. As the following theorem shows, any Poisson manifold
$P$ can be obtained in the above way:

\begin{theo}[Karasev, Weinstein \cite{SympReal,Weinstein}] 
 Every Poisson manifold has a surjective, submersive symplectic
 realization.
\label{realization}
\end{theo}

In a sense dual to this observation is the following embedding
theorem for a presymplectic manifold:
\begin{theo}[Gotay \cite{Gotay}]
\label{Gotay} Every presymplectic manifold $(M,\o)$
can be coisotropically embedded into a symplectic manifold. This
extension of $M$ is unique up to a local symplectomorphism in a
neighborhood of $M$.
\end{theo}

Later in this paper (Sec.~\ref{sec:LeafSymp}) we will address the
question as to whether there is an analogous, rather than dual, result
for Poisson manifolds, namely whether any Poisson manifold $P$ can be
embedded into a symplectic manifold $M$ such that the embedding
becomes symplectic upon restriction to any leaf of $P$.

We may also compare this with the embeddings considered in the
previous section (cf.\ in particular Definition \ref{coisoPoi}): The
embedding $\iota$ of a second class submanifold $C$ into a Poisson
manifold cannot be Poisson. Indeed in order to be able to define some
$\Pi_C \subset \Gamma(\Lambda^2 TC)$
such that $\iota$ is Poisson ($\iota_* \Pi_C = \Pi_M$), it is
necessary and sufficient that $\Pi_M^\sharp(T^*M)|_C \subset TC$,
which is easily recognized as the condition (\ref{ideal}), and which
obviously can be defined only for $\dim C \ge \rank \Pi_M^\sharp$
(while a for a second class submanifold always $\dim C \le \rank
\Pi_M^\sharp$, equality holding for $C$ being a symplectic
leaf). Nevertheless $C$ inherits a natural (nondegenerate) Poisson
structure as a submanifold of a symplectic leaf. It may be quite
interesting to reconsider (and possibly generalize) the (embedding)
maps discussed in Definition \ref{coisoPoi} and
Sec.~\ref{sec:LeafSymp} within the more general and unifying framework
of Dirac structures (cf.\ footnote \ref{footnote3}).

\medskip

In the context of constrained systems $(M,\omega)$ with a closed
constraint algebra (\ref{consalg}) there is, under appropriate
regularity conditions, also another canonical Poisson manifold $Q$
associated to it: Define an equivalence
relation on $M$ by calling two points equivalent, if they lie on the
same orbit generated by the Hamiltonian vector fields $v^\alpha = \{
\cdot , \Phi^\alpha \}$ of the constraints. Assume that $Q:=
M/\!\!\sim$ is a differentiable manifold (for a case where this
condition is violated \cf e.g.\ Example 2 below) and denote the
respective projection map from $M$ to $Q$ by $\pi$. Then $Q$ may be
equipped with a Poisson bracket by $\pi^* \{ f,g \}_Q := \{ \pi^* f,
\pi^* g \}_M$ for all $f,g \in C^\infty(Q)$.  This indeed provides a
well-defined bracket on $Q$ since due to the Jacobi identity for the
Poisson bracket on $M$ the right-hand side is a function in the kernel
of all the $v^\alpha$ and thus it can be written as the pull back of a
function on $Q$. With this choice for the bracket, $(Q,\Pi_Q)$ is a
Poisson quotient of $M$, \ie the projection map $\pi\colon M\to Q$ is
Poisson.

The two manifolds $Q$ and $P$ are not unrelated certainly. In fact we
have
\begin{prop} 
 With the assumptions above the two Poisson manifolds $Q$ and $P$ form
 a symplectic dual pair with respect to the original symplectic
 manifold $M$, in which the Poisson map $\pi\colon M \to Q$ is
 complete.\label{prop1}
\end{prop}

\begin{defi}[\cite{SW}] Two Poisson manifolds $(P,\Pi_P)$ and
$(Q,\Pi_Q)$ form a {\em symplectic dual pair\/} with respect to a
symplectic manifold $(M,\omega)$ if there are two Poisson maps
$\phi\colon (M,\omega)\to (P,\Pi_P)$, $\pi\colon (M,\omega)\to
(Q,\Pi_Q)$ with symplectically orthogonal fibers.
\begin{equation} \mbox{\parbox[c]{5cm}{\setlength{\unitlength}{1.5mm}
\begin{picture}(30,30) \put(15,25){\makebox(0,0)[t]{$(M,\omega)$}}
\put(0,8){\makebox(0,0)[t]{$(P,\Pi_P)$}}
\put(30,8){\makebox(0,0)[t]{$(Q,\Pi_Q)$}}
\put(8,15){\makebox(0,0)[t]{$\phi$}}
\put(21,15){\makebox(0,0)[t]{$\pi$}} \put(14,22){\vector(-1,-1){13}}
\put(16,22){\vector(1,-1){13}}
\end{picture}}} \label{diagram1}
\end{equation}
\end{defi}

\begin{defi}
 A Poisson map $\pi\colon M\to Q$ is {\em complete}, if the
 Hamiltonian vector field $X_{\pi^*h}=\{\cdot,\pi^*h\}$ generated by
 $\pi^*h$ in $M$ is complete for any function $h$ of compact support
 on $Q$.
\end{defi}

\begin{proof}
To prove the first part of the above Proposition it is only necessary
to note that the fibers are symplectically orthogonal iff the pull
back of any function on $P$ (\ie a function of the constraints)
Poisson commutes with the pull back of any function on $Q =
M/\!\!\sim$, \ie a function $f$ satisfying $v^{\alpha}f=0$ for all
$1\leq\alpha\leq d$.  Any function $g\in\phi^* C^{\infty}(P)$ can be
expressed as a function of the $\Phi^{\alpha}$ only, for which we have
$\{\Phi^{\alpha},f\}= -v^{\alpha}f=0$, proving $\{g,f\}=0$ for any
$f\in\pi^* C^{\infty}(Q)$, $g\in\phi^* C^{\infty}(P)$ and so
symplectic orthogonality of the $\pi$- and $\phi$-fibers.

The completeness of the projection map $\pi$ may be shown by adapting
the proof of Proposition 6.6 in \cite{SW}: Let $h$ be a function of
compact support on $Q$ which implies that $X_h=\{\cdot,h\}$ has a
complete integral curve through some given point $q\in Q$. We now
assume that the integral curve of $X_{\pi^*h}$ through any point
$m\in\pi^{-1}(q)$ is not complete, \ie there is (without loss of
generality) a maximal parameter $t_+$ beyond which the curve cannot be
defined. Through the projection to $Q$ by $\pi$ of the final point
$m_+$ belonging to $t_+$, there is, however, an extension of the
integral curve which can be lifted to $M$ to a curve through some
point $m'\in\pi^{-1}(\pi(m_+))$ (not necessarily identical to
$m_+$). Using the action generated by the constraints $\Phi^{\alpha}$
on $M$, by which we defined the equivalence relation $\sim$, we can
construct local symplectomorphisms $s$ from a neighborhood of $m'$ to
a neighborhood of $m_+$ such that $s(m')=m_+$ (e.g., by following
integral curves of the $v^{\alpha}$ in a chosen ordering, where the
parameter lengths of all the integral curves are determined so as to
fulfill $s(m')=m_+$ and are the same for all points in the
neighborhood of $m'$). Since $X_{\pi^*h}$ is by definition invariant
under the flow, the local symplectomorphism transports a piece of its
integral curve through $m'$ to a piece of the integral curve through
$m_+$. This is a contradiction to our assumption that the original
integral curve through $m$ is inextendible beyond $m_+$ which proofs
completeness of the map $\pi$.
\end{proof}

The orthogonality in the above proof could have been infered also 
from the results of Sec.\ \ref{Prelim}, where the span of the
vector fields $v^\alpha$ generating the orbits (the fibers of the
map $\pi$) was identified with $(TC)^\perp$, the symplectically
orthogonal to $TC$, which in turn is the tangent bundle to a 
(generalized) ``constraint surface'' $C$ defined by the preimage
of {\em some\/} point in $P$ (the fibers of the map $\phi$). 
Note in this context that orthogonality of the fibers does not
imply that they are transversal to one another (this is only the
case for second class constraints). In particular, for first
class constraints the fibers of $\pi$ are even submanifolds
of the fibers of $\phi$.

In the rest of this section we want to clarify the relation between
the factor space $Q$ and the reduced phase space $R$. For instance if
$P$ is symplectic, so that the constraints $\Phi^\alpha$ are second
class, $R$ coincides just with $(C,\omega_C)$ in (\ref{P}). We then
will show that, under some rather mild conditions specified below, $R$
and $Q$ are symplectomorphic. (Note that $P$ symplectic implies $Q$
symplectic because their pre-images in the symplectic manifold $M$ are
symplectically orthogonal due to Prop.\ \ref{prop1}.)  Thus in this
case we can trade in the standard procedure of reducing the
corresponding second class constrained system on $M$ by restriction to
the constraint surface $C$ for taking the factor space of $M$ with
respect to the flow generated by the {\em second class\/}
constraints. This second approach is in its spirit more closely
related to the (in physical systems generically better understood)
reduction procedure used for constrained systems with purely first
class constraints. In fact, as we will show, $M$ may be regarded as a
constraint surface in a higher dimensional (extended) symplectic
manifold $\widetilde M$ characterized by first class constraints, the
orbits of which on $M \hookrightarrow \widetilde M$ coincide with the
original orbits of $v^\alpha$ on $M$.

In the following we will first focus on this case of a symplectic
target $(P,\Pi_P)$ (corresponding to second class constraints) before
we then consider the general case.

\subsection{Second class constraints}
\label{sec:Second}

By assumption, $(P,\Pi_P)$ is symplectic (we denote the symplectic
form inverse to $\Pi_P$ by $\O_P$ in the following).  Then any Poisson
map $\phi\colon M\to P$ is a submersion \cite{SW}. If in addition
$\phi$ is {\em complete}, we have

\begin{lemma}\label{bundle} Let $\phi$ be a complete Poisson map from
a Poisson manifold $M$ to a symplectic manifold $(P,\Omega_P)$. Then
$M$ is a fiber bundle over $P$ with projection map $\phi$ which is
endowed with a natural flat connection.
\end{lemma}

The proof \cite{SW} proceeds by showing that any vector $v\in TP$ can
be lifted horizontally to a vector in $TM$ by pulling back the
covector $v\lrcorner\Omega_P$. In this way one obtains a connection
which is flat since the Lie bracket of two horizontal vector fields is
again horizontal. To see this, one uses canonical coordinates on $P$
and completeness of the map $\phi$.

In our situation of regular second class constraints, we have

\begin{theo}\label{dual} 
Let $\Phi^{\alpha}$, $1\leq\alpha\leq d$ be regular constraint
functions with nonempty constraint surface $C=\phi^{-1}(0)$ on a
connected symplectic manifold $(M,\o)$ with a complete Poisson map
$\phi\colon (M,\o)\to (P,\Pi_P),x\mapsto\Phi^{\alpha}(x)$ such that
the Poisson manifold $(P,\Pi_P)$ is symplectic.

If $Q=M/\!\!\sim$, which is symplectically dual to $(P,\Pi_P)$ with
respect to  $(M,\omega)$, is a manifold, then it is covered by the
constraint surface $(C,\omega_C)$. This is always the case if the
holonomy  group of the flat connection of Lemma  \ref{bundle}
is finite; if it is trivial, the covering by $C$ is a symplectomorphism.
\end{theo}

\begin{proof}
By construction $C$, defined by $\Phi^{\alpha}=0$ for all $\alpha$, is
embedded as a submanifold into $M$, $\iota \colon C \to M$, and $\o_C
= \iota^* \o$. Now define an equivalence relation for points on $C$
by calling two points on $C$ equivalent if they are connected by a
constraint orbit in $M$. Taking the corresponding factor space yields
$Q$ as a topological space. Here it is essential that any orbit
$\CO_p$ of the Hamiltonian vector fields generated by $\Phi^{\alpha}$
through a point $p\in M$ intersects the constraint surface $C$ at
least once.

This is indeed the case due to completeness of the map $\phi$: if 
$p\not\in C$, the point $p$ is mapped to $0\not=\phi(p)\in P$, which
can be connected to $0\in P$ along trajectories of Hamiltonian
vector fields, chosen to be of compact support, in $P$ since $P$ is
symplectic. These trajectories can be carried to $M$ by pulling back
their Hamiltonian functions with $\phi$. Due to the completeness of
$\phi$, they can be extended to arbitrary parameters such that we
reach a point in the pre-image of $0\in P$. This point lies on the
constraint surface $C$.

In general, there will be more than one intersection point.  Thanks to
Lemma \ref{bundle} the bundle $\phi\colon M\to P$ with typical fiber
$C=\phi^{-1}(0)$ comes equipped with a flat connection whose holonomy
provides an action of the fundamental group $\pi_1(P)$ on $C$. The set
of all points in the intersection of $C$ with an orbit $\CO$ through a
point $p\in C$ is given by the orbit through $p$ of the action of
$\pi_1(P)$. The number of copies of $p$ obtained in such a way is
given by the number $k$ (maybe infinite) of elements in the factor
space $\pi_1(P)/F_p$, where $F_p$ is the isotropy subgroup of
$\pi_1(P)$ in $p$. Since $\pi_1(P)$ is discrete, this number is
constant on any connected component of $C$. Factoring $C$ with the
action of $\pi_1(P)$ yields a $k$-fold covering $C\to Q$, where $Q$ is
obtained by the projection map $\pi\colon M\to Q$ along the orbits
$\CO$ as in Proposition \ref{prop1}. If $k$ is finite, the action of
$\pi_1(P)$ is properly discontinuous and $Q$ is a manifold (see
Example 2 for a case with an infinite holonomy group).  If $k=0$
(trivial holonomy), $C \cong Q$ topologically.

We next show that $\o_C$ factors through to $Q$ (under the assumption
that $Q$ is a manifold). Hamiltonian vector fields generated by
$\Phi^{\alpha}$ yield a {\em 
symplectic\/} map between neighborhoods of the surfaces
$\phi(x)=\phi_0$ and $\phi(x)=\phi_1$ for any constants $\phi_0$ and
$\phi_1$. This also holds true if $\phi_0=\phi_1=0$, in which case we
have the map identifying points on $C$ to result in $Q$ used
above. Consequently, the symplectic structure $\o_C$ on $C$ can be
projected down to yield the symplectic manifold $(Q,\O_Q)$.

It remains to show that the map $\pi\colon (M,\o)\to (Q,\O_Q)$ is
indeed Poisson. (The rest, such as symplectic orthogonality of the
$\pi$- and $\phi$-fibers, follows from (or as in the proof of)
Proposition \ref{prop1}.) For this purpose we introduce some further
notation: Decompose the tangent and cotangent space at $x\in C \subset
M$ according to the splitting induced by $C$ and the orbit
$\CO_x$. More explicitly, $T_xM \cong T_xC \oplus T_x\CO$ and $T^*_xM
= \Ann_M(T_x \CO) \oplus \Ann_M(T_x C) \cong T^*_xC \oplus T^*_x\CO$,
where $\cong$ denotes canonical isomorphisms, used to identify
respective vector spaces.\footnote{We could have replaced $T_x \CO$ by
$(T_x C)^\perp$, cf.~Sec.~\ref{Prelim}.}  Denote by $\iota_i$ and
$p_i$, $i=1,2$, the respective embeddings and projections to the first
and second factor in $T_xM$; thus e.g.\ $\iota_1 \colon T_xC \to
T_xM$.  Then $\pi_i := \iota_i \circ p_i $ is the corresponding
projection operator inside $T_xM$. Proceed likewise for $T_x^*M$ with
bars on the respective maps; thus, e.g., $\bar \iota_1 \colon T^*_xC
\to T^*_xM$ etc.

In this notation, the condition defining $\o_C$, $\o_C = \iota^* \o$,
takes the form $\omega_C^\sharp=\bar p_1 \o^\sharp  \iota_1$,
while  orthogonality of the fibers becomes
$p_2 \Pi^\sharp \bar \pi_1 = 0 = p_1 \Pi^\sharp \bar \pi_2$.
{}From this we need to prove that $\pi$ as defined above is Poisson,
\ie  that $p_1 \Pi^\sharp \bar \iota_1 = \Pi_C^\sharp$.  This indeed
follows since  
\[ 
 p_1 \Pi^\sharp \bar \iota_1 \omega_C^\sharp
 =  p_1 \Pi^\sharp \bar \iota_1 \bar p_1 \omega^\sharp \iota_1 =
 p_1 \Pi^\sharp (\bar \pi_1 + \bar \pi_2) \omega^\sharp \iota_1
 = -\id_{T_xC}
\]
\end{proof}

\noindent{\bf Example 1:} (generalized from \cite{ASS}) Let
$M:=\R^2\backslash\{(0,0)\}$ be the punctured plane with canonically
conjugate coordinates $(q,p)$ and define on $M$ the constraints \[
\Phi^1_n:=\sqrt{q^2+p^2}\,T_n\left(q(q^2+p^2)^{-\frac{1}{2}}\right)-1
\quad,\quad \Phi^2_n:=
p\,U_{n-1}\left(q(q^2+p^2)^{-\frac{1}{2}}\right) \] for a fixed
$n\in\N$ using the Chebyshev polynomials $T_n$ and $U_n$ of the first
and second kind. With $T_n(\cos\theta)=\cos n\theta$ and
$U_n(\cos\theta)= \sin[(n+1)\theta]/\sin\theta$, one can see that
$\{\Phi^1_n,\Phi^2_n\}=n$, and thus
$P=\phi(M)=\R^2\backslash\{(-1,0)\}$ is a symplectic manifold.

 In polar coordinates defined by $q=r\cos\phi$, $p=r\sin\phi$ it
 follows easily that the constraint surface $C=\{(\cos
 (k\pi/n),\sin (k\pi/n)):0\leq k\leq n\}$ consists of $n$
 points which are all connected by the single orbit generated by the
 constraints in $M$. The map $C\to Q$ is an $n$-fold covering.

 \medskip

\noindent{\bf Example 2:} 
Let $M:=T^2\times\R^2\ni
(q_1,p_1;q_2,p_2)$ with constraints $\Phi^1:=q_1+\omega
p_1+q_2$ and $\Phi^2:=p_2$ (the first constraint function
is taken modulo 1 such that $P=S^1\times\R$).

If $\omega=0$, the constraint surface is homeomorphic to $S^1\times\R$
where the $S^1$-factor comes from the unaffected coordinate $p_1$, and
the $\R$-factor is the regular curve (a helix) $q_1=-q_2$ (mod 1)
lying on the cylinder $S^1\times\R\subset M$ with coordinates $q_1$,
$q_2$. The flow generated by the second constraint intersects the
constraint surface of the first constraint in an infinite number of
points of the form $(q_1,p_1,-q_1+m,0)$ for $m\in\Z$, and vice
versa. Since the corresponding identifications form a discrete set of
translations in the $q_2$-direction of constant shift, the factor
space $Q$ is a manifold.

The situation is unchanged if $\omega$ is nonvanishing but
rational. But if $\omega$ is irrational, the projection of the
constraint surface to the cylinder appearing above will be
dense. There are still infinitely many intersection points, and this
time the factor space is not a differentiable manifold. This also
follows from the fact that the orbits generated by the constraints are
dense in a suitable subset of $M$, and so $Q=M/\!\!\sim$ cannot be a
manifold.

\bigskip

Theorem \ref{dual} can be used to define a natural presymplectic form
on $M$:

\begin{cor}[\cite{ASS}]\label{ASS} Under the assumptions of Theorem
\ref{dual} the symplectic manifold $(M,\o)$ carries a presymplectic
form \begin{equation}\label{baro} \bar{\omega}:=\omega-\phi^*\Omega_P
\end{equation} whose kernel is tangent to the orbits generated by the
$\Phi^{\alpha}$.
\end{cor}

\begin{proof} By definition $\bar{\o}$ is closed. From Theorem
\ref{dual}, in particular the fact that the fibers of the dual pair
are symplectically orthogonal, it follows that $\bar{\o}$ vanishes
when applied to vector fields tangential to the orbits $\CO$. This can
also be seen directly: By definition, the Hamiltonian vector field
$X_{\Phi^{\alpha}}$ generated by $\Phi^{\alpha}$ satisfies
$X_{\Phi^{\alpha}}\lrcorner\omega= \md\Phi^{\alpha}$.
Furthermore, by definition
of $\Omega_P$, we have
\[ 
 X_{\Phi^{\alpha}}\lrcorner\phi^*\Omega_P=
 \left(\Omega_P\right)_{\beta\gamma} X_{\Phi^{\alpha}}
 (\Phi^{\beta})
 \md\Phi^{\gamma}= -(\Pi_P^{-1})_{\beta\gamma}
 \Pi_P^{\beta\alpha}\md\Phi^{\gamma}=
 \md\Phi^{\alpha}
\] 
so that $X_{\Phi^{\alpha}}\lrcorner\bar{\omega}=0$.  \end{proof}

It is clear from the proof of Theorem \ref{dual} that we have a
trivial covering by $C$ if $\pi_1(P)=0$:

\begin{cor}\label{simplyconn} If $P$ is simply connected, then
$(P,\Pi_P)$ and $(C,\o_C)$ form a symplectic dual pair with respect to
$(M,\o)$. 
\end{cor}

This is always the case if $M$ is simply connected, as we will now
demonstrate. In Theorem \ref{dual} and its proof we made use of the
fact that $M$ may be considered as a fiber bundle over $P$ with
typical fiber $C=\phi^{-1}(0)$, the constraint map $\phi$ playing the
role of the projection from $M$ to $P$. However, if $Q$ is a manifold,
$(Q,\Pi_Q)$ is symplectic and, according to Proposition \ref{prop1},
the respective projection map $\pi$ is a {\em complete\/} Poisson
map. We thus may apply Lemma \ref{bundle} also to this context and
conclude as in the proof of Theorem \ref{dual} that the orbit
$(\CO,\o_\CO)$ (with symplectic form $\o_\CO$ induced on it by $\o$)
covers $(P,\O_P)$. Moreover, this covering map is a symplectomorphism
if the holonomy of the respective flat connection is trivial, which in
particular is always the case if $\pi_1(Q)=0$.

\begin{prop}
If $M$ is simply connected (and connected), also $P$ and $C$ are
simply connected and connected. \label{connected}
\end{prop}

\begin{proof} 
 For the fiber bundle $\CO \hookrightarrow M\to Q$ we have the exact
 homotopy sequence (\cf e.g.\ \cite{BottTu}) \[
 \cdots\to\pi_1(M)\to\pi_1(Q)\to\pi_0(\CO) \to\cdots \] Since by
 assumption $\pi_1(M) =0$ and by definition $\pi_0(\CO)=0$, it follows
 that $\pi_1(Q)$ vanishes. According to the considerations above, this
 implies that the fiber bundle $\CO \hookrightarrow M\to Q$ is trivial
 and that its fiber is isomorphic to $P$ (here we use the holonomy of
 Lemma~\ref{bundle} for which the completeness of the map $\phi\colon
 M\to P$ is important); thus $M \cong Q \times P$.  This in turn
 yields $\pi_1(M) = \pi_1(Q) \oplus \pi_1(P)$, resulting in
 $\pi_1(P)=0$. Thus $Q \cong C$, $\pi_1(C)=0$, and with $M$ being
 connected, finally also $P$ and $Q$ (or $C$) are connected.
\end{proof}

This result can be helpful because $M$ is the original symplectic
manifold and may thus be easier accessible than $P$ in particular
cases.  We also note that $P$ being simply connected is only a
sufficient condition for $Q\cong C$. The essential object which
determines $Q$ is the holonomy of the flat connection of Lemma
\ref{bundle}, which certainly can be trivial even if $P$ is not simply
connected. In other words, it is the behavior of the
$v^{\alpha}$-orbits in $M$ which determines whether the reduced phase
space $C$ itself and $P$ are symplectically dual with respect to $M$
(and consequently then $C$ may be obtained as factor space by
projection $\pi$ along the orbits).

\medskip

\noindent{\bf Example 3:} Let $M=T^*(S^1 \times \R)$ with canonically
conjugate coordinates $(x,y; p_x,p_y)$, where $x \equiv x+1$, and with
constraints $\Phi^1 := x$, $\Phi^2 := p_x$. Then $P=T^*S^1$ is not
simply connected, but still $R\equiv C=Q=T^*\R$.

\medskip

According to Proposition \ref{connected}, $M$ being simply connected
implies $Q\cong C$ and also that $C$ is connected.  In Example
1, $M$ was not simply connected, $C$ was not connected and $C$ was
a nontrivial covering of $Q$, \ie we had Gribov copies (a Gribov copy
of a point on the constraint surface is a different point on the
constraint surface which lies on the same orbit generated by the
constraints \cite{ASS,Gribov}).  As the following example
demonstrates, this can also happen if $C$ is connected (but $M$ still
is not simply connected):

\medskip

\noindent{\bf Example 4:} Let $M$ be the 4-torus with canonically
conjugate coordinates $(x,p_x;y,p_y)\in T^2\times T^2$ which are
identified modulo 1. The constraints are $\Phi^1:=nx+y$ and
$\Phi^2:=p_x$ with $1<n\in\N$ such that $\{\Phi^1,\Phi^2\}=n$ and the
constraint surface $C\cong T^2$ characterized by $(x,0,-nx,p_y)$ is
connected.

The Hamiltonian vector fields generated by the constraints are $v^1
=-n \md/\md p_x -\md/\md p_y$ and $v^2=\md/\md x$.  Their orbit ${\cal
O}$ parameterized by $t_1$, $t_2$ through the point $(x_0, 0, -nx_0,
p_{y,0})$ is
\[ \exp(t_1v^1+t_2 v^2)(x_0,0,-nx_0,p_{y,0})=
(x_0+t_2,-nt_1,-nx_0,p_{y,0}-t_1)\,.
\] It intersects $C$ if and only if $t_1=m_1/n$, $t_2=m_2/n$ with
$m_1,m_2\in\Z$. On the other hand, points on ${\cal O}\cong T^2$
coincide if $t_1$ or $t_2$ are changed by an integer. Therefore, the
intersection ${\cal O}\cap C$ consists of $n^2$ points, which are
obtained with $1\leq m_1,m_2\leq n$.  

Thus, $C$ is connected, $M$ not simply connected and we have $n^2$
Gribov copies.  We note that the map $\phi|_{\cal O}\colon
(x_0+t_2,-nt_1,-nx_0,p_{y,0}-t_1)\mapsto (nt_2,-nt_1)$ 
also is an $n^2$-fold covering of $P\cong T^2$.

\medskip

So the connectedness of the constraint surface $C$ (which coincides
with the reduced phase space $R$ of physical interest) alone is not
sufficient for the absence of Gribov copies, \ie to guarantee together
with completeness of $\phi$ that any orbit ${\cal O}$ generated by the
constraints intersects $C$ in precisely one point. However, in the
above example, $P$ is not a subset of $\R^2$, which would usually be
the case in physical models (so, e.g., $\Phi^1 \not \in C^\infty(M)
\equiv C^\infty(M,\R)$ and, correspondingly, the orbit generating
vector fields $v^1$ and $v^2$ are only {\em locally\/}
Hamiltonian). It would be interesting to clarify whether the condition
$\pi_0(C)=0$ is sufficient if $P$ is a subset of $\R^d$, $d=\dim P$.

We now summarize our findings in a diagram:

\begin{equation} \mbox{\parbox[c]{6cm}{
\begin{picture}(60,40)(35,37)
\put(60,70){$(M,\omega)$}
\put(70,68){\vector(3,-2){30}}
\put(66,68){\vector(0,-1){20}}
\put(31,49){\rotatebox{130}{\oval(3,3)[t]}}
\put(32,48){\vector(3,2){30}}
\put(22,42){$(R,\omega_R)\hspace{4mm}
\stackrel{\cong}{\longrightarrow}\hspace{4mm} 
(Q=M/\!\!\sim,\O_Q)\hspace{17mm}(P,\O_P)$}
\put(44,60){$\iota$}
\put(68,57){$\pi$}
\put(86,60){$\phi$}
\end{picture}}} \label{diagr:second}
\end{equation}

In the present subsection we considered the case of a symplectic
target $(P,\O_P)$ (the general case of a Poisson manifold will be
considered in the subsequent subsection). For simplicity we take the
original unconstrained phase space $(M,\o)$ to be connected. We now
require the orbit space $Q=M/\!\!\sim$ to be a differentiable manifold
(the orbits are generated by the constraints $\Phi^\alpha$). The right
hand side of the diagram then follows from Proposition~1, yielding
$\pi$ to be a complete Poisson map. When $P$ is symplectic, the
constraints are second class and the reduced phase space $(R,\o_R)$
coincides with the constraint surface $(C,\o_C)$ as embedded in
$M$. Thus with (\ref{P}) we obtain the embedding of $(R,\o_R)$ into
$M$ in the diagram. Requiring that the constraint map $\phi$ is
complete, too, we showed in the present subsection that $(R,\o_R)$
covers the Poisson quotient $Q$ via the map $\pi\circ\iota$. Moreover,
if (as a sufficient but not necessary condition) $\pi_1(P) = 0$ (which
in turn follows e.g.\ if $M$ is simply connected), this covering is an
isomorphism (`no Gribov copies'). (In \cite{ASS} also another
potential source for the map $R\to Q$ not to be an isomorphism was
mentioned and backed up by an explicit example, namely systems where
the map $\phi$ restricted to an orbit is not surjective.  This
qualitatively different case of a Gribov problem (corresponding to
`gauge orbits' not intersecting the `gauge fixing surface' $C$) is
excluded by the requirement that $\phi$ should be complete.)

In this favorable case of an isomorphism we thus obtain the reduced
phase space $(R,\o_R)$ not only by restriction to the constraint
surface $C \subset M$ (standard procedure for reducing second class
constraints), but instead may consider the factor space $Q$ of $M$
with respect to the orbits generated by the constraints. This is more
closely related to the standard and generally more preferred treatment
of first class constraints.

In fact, the analogy with first class constraints can be made even
more precise: According to Corollary
\ref{ASS}, $M$ may be equipped with a presymplectic form $\bar
\omega$, the kernel of which coincides with the orbits to be factored
out. Together with Corollary \ref{simplyconn} we have

\medskip

\noindent {\bf Corollary 2'} \ {\it If the Poisson map $\phi$ of the
constraints is complete and the target $P$ symplectic and simply
connected,  
there exists a symplectomorphism from $(M,\o)$ to
$(R,\omega_R) \times (P,\O_P)$ with canonical projections to both
factors.  The reduced phase space may then be obtained by projecting
to the first factor or, equivalently, by taking the factor space of
$(M,\bar \o)$ with respect to the kernel of the presymplectic form
$\bar \o$ defined in (\ref{baro}).}

\medskip

The presymplectic manifold $(M,\bar \omega)$ can then in turn be
embedded coisotropically into some higher dimensional symplectic
manifold $(\widetilde M, \widetilde \omega)$. This extended phase space is
equipped with first class constraints such that $M$ is the level zero
set of the constraints and $\bar \omega$ the pull back of $\widetilde
\omega$ with respect to the corresponding embedding map. In this way,
$(R,\o_R)$ can be obtained by standard reduction of the extended phase
space $(\widetilde M,\tom)$ constrained by {\em first class\/}
constraints.  In section \ref{subsecAppl} we will provide two explicit
possibilities for constructing such a constrained extended phase space
$\widetilde M$, in one of which the constraints are even Poisson commuting
(abelian).  Within this reformulation, the above mentioned orbits of
the (originally second class) constraints $\Phi^\alpha$, $\alpha = 1,
\ldots, d$, become just the standard gauge orbits of the extended
system (generated by $d$ first class constraints) and the original
constraint surface $C$ corresponds to one possible choice of a gauge
(which exists globally due to the absence of a Gribov problem---now in
the standard use of the terminology---as a consequence of the
assumptions specified above).

Let us add a remark concerning the arrows in diagram
(\ref{diagr:second}): As remarked already before, they are morphisms
of respective categories. So $\iota$ as well as the covering map
$\iota \circ \pi$  are symplectic maps and $\pi$, $\phi$ are Poisson. 
Note that neither of the last two maps
is symplectic, since $\pi^* \O_Q = \o - \phi^* \O_P \neq
\o$. Similarly, $\iota$ is also not Poisson (except for $C=M$), since
a Poisson map between symplectic manifolds is always a submersion
\cite{SW}. Only the covering map is symplectic {\em and\/} Poisson. 
Likewise remarks apply to the diagrams below.

We finally remark that it is a direct consequence of Proposition
\ref{connected} that, provided $M$ is connected and simply connected,
so are $(R,\o_R)$ and $\bar{P}:=(P,-\Pi_P)\cong
(P,-\Omega_P)$. Moreover, in this case $M\cong R\times P$ and
the maps $\pi$ and $\phi$, which then are just projections to the
first and second factor, respectively, have constant rank. This
demonstrates that $R$ and $\bar{P}$ are {\em Morita equivalent\/} (cf
\cite{SW} for the definition) with respect to $(M,\o)$.

\subsection{The general case}

Let us first consider the other extreme case opposite to the one of
Sec.~\ref{sec:Second} and cast the standard reduction of a system
of first class constraints into a diagram similar to the one at the
end of the previous subsection. This yields the following
(commutative) diagram:

\begin{equation} \mbox{\parbox[c]{7cm}{
\begin{picture}(70,39)(40,34)
\put(30,70){$(C=\phi^{-1}(0),\omega_C)\
\stackrel{\iota}{\longrightarrow}\ (M,\omega)$}
\put(64,71.8){\rotatebox{90}{\oval(1.5,1.5)[t]}}
\put(45,68){\vector(0,-1){20}}
\put(75,68){\vector(0,-1){20}}
\put(79,68){\vector(3,-2){28}}
\put(38,42){$(R,\omega_R)\hspace{2mm}\hookrightarrow \hspace{2mm}
(Q=M/\!\!\sim,\Pi_Q)\hspace{15mm}(P,\Pi_P)$}
\put(47,57){$\pi|_C$}
\put(77,57){$\pi$}
\put(98,57){$\phi$}
\end{picture}}}\label{diagr:first}
\end{equation}

The right hand side is again taken from Prop.\ \ref{prop1} (the
assumption from there that $Q=M/\!\!\sim$ is a differentiable manifold
is understood to hold here too). By definition the constraint surface
$C$ is the preimage of zero with respect to the constraint map
$\phi$. Restricting the projection map $\pi$ to $C$, which is
equivalent to factoring out the flow generated by the kernel of
$\omega_C$, yields the standard reduced phase space $R$.

\begin{prop} \label{firstclassprop}
 Let $(R,\omega_R)$ be the reduced phase space of a first class
 constrained system with the assumptions of Prop.\ \ref{prop1}. Then
 $R$ is a symplectic leaf in the orbit space
 $(\mbox{$Q=M/\!\!\sim,$}\, \Pi_Q)$ which is obtained by factoring out
 the constraint orbits.
\end{prop}

\begin{proof}
 Note first that there is a natural embedding of $R$ as a subset of
 $Q$, since the flow of the constraints does not leave $C$ being the
 preimage of a symplectic leaf (namely the origin) in $(P,\Pi_P)$.

 To prove that the embedding is as a symplectic leaf, we first note
 that $\Pi_Q^{\sharp}(\Ann_Q(TR))=0$  if $R$ is the reduced phase
 space. This follows from the fact that $\Ann_Q(TR)$ is spanned by
 total derivatives of the constraints which on $R$ Poisson commute
 with each other  as well as with elements of
 $T^*Q|_R/\Ann_Q(TR)$   since this space is spanned by total
 derivatives of physical observables which by definition commute on
 $R$ with the constraints. This shows that we can define a Poisson
 structure $\Pi_R$ on $R$ using (\ref{PoissStruc}):
\[
 \{f,g\}_R= \Pi_R(\md f,\md g)= \Pi_Q((\md f)_Q,(\md g)_Q)|_R
\]
 using the isomorphism $i\colon T^*R\to T^*Q|_R/\Ann_Q(TR),
 \alpha\mapsto \alpha_Q+\Ann_Q(TR)$. If this Poisson structure is
 nondegenerate, $R$ is an open subset of a symplectic leaf in $Q$.

 We proceed by showing that the pull back of $\{f,g\}_R$ under
 $\pi|_C$ coincides with the bracket computed using the pull back of
 the symplectic structure on $M$. This will imply that $\Pi_R$
 coincides with the structure of $R$ as the reduced phase space which
 by definition is nondegenerate.

 In a first step we use the fact that $\pi$ is Poisson and obtain
\[
 \pi|_C^*\{f,g\}_R=\Pi_M(\pi^*(\md f)_Q,\pi^*(\md g)_Q)|_C\,.
\]
 Here we need the image under the map $i$ of an exact 1-form $\md
 f$. More precisely, we want to show that $\alpha_Q$ can be chosen to
 be exact if $\alpha=\md f$ is exact. To construct a function $f_Q$ on
 $Q$ with $\alpha_Q=\md f_Q$, we choose a tubular neighborhood of $R$
 in $Q$, cover it with open subsets in which we can use local
 coordinates of $R$ together with transversal coordinates, and use a
 partition of unity subordinate to these neighborhoods. First we can
 extend $f$ to a function $f_{{\cal U}}$ on any local neighborhood
 ${\cal U}$ by requiring that $f_{{\cal U}}$ does not depend on the
 transversal coordinates. Using the partition of unity, we arrive at a
 smooth function which is defined on the full tubular neighborhood of
 $R$ and which equals $f$ when pulled back to $R$. Multiplying this
 function with a smooth function which vanishes outside the tubular
 neighborhood and equals one on $R$ defines a smooth function $f_Q$
 defined on $Q$. Because $f_Q|_R=f$, we have $v(f_Q)=v(f)$ for any
 vector $v$ tangential to $R$ such that we can choose $(\md f)_Q=\md
 f_Q$ as representative in $T^*Q|_R/\Ann_Q(TR)$. We use this relation
 to compute the pull back of the Poisson bracket
\[
 \pi|_C^*\{f,g\}_R= \Pi_M(\md\pi^* f_Q,\md\pi^* g_Q)|_C=
 \{\pi^*f_Q,\pi^* g_Q\}|_C\,.
\]
 Owing to the construction of $f_Q$ above, $\pi^*f_Q$ is a function on
 $M$ which is constant along constraint orbits and whose values on the
 orbits through $C$ coincide with $f$.  This shows that the pull back
 to $C$ of the symplectic structure on $R$ as a leaf in $Q$ coincides
 with the symplectic structure of $M$ restricted to $C$.

 Completeness of the map $\pi$ implies that $R$ is a symplectic leaf
 in $Q$, not just an open subset of a leaf: Assume that $R$ is
 contained in but not identical to a leaf $L$ of $Q$ and choose a
 point $r\not\in R$ in the boundary of $R$ in $L$. We can choose a
 function $h$ on $Q$ which is supported only on some neighborhood of
 $r$ and which generates a Hamiltonian vector field $X_h$ tangential
 to $L$. Using the trajectories of $X_h$, we can connect the point
 $r\not\in R$ to a point $r_0\in R$. Due to completeness, the pull
 back $\pi^*h$ generates a complete Hamiltonian vector field on
 $M$. Furthermore, it is tangential to $C$ because $X_h$ was chosen to
 be tangential to $L$. Thus, there are points in $\pi^{-1}(r)$ and
 $\pi^{-1}(r_0)$ which both lie in $C\subset M$ and so are projected
 to $R$ under $\pi$. Therefore, $r\in R$ contradicting our assumption
 $R\not=L$.
\end{proof}

Thus for the reduction process in a system of first class constraints
{\em forming a closed constraint algebra\/} we may exchange the order
of restriction to a submanifold and taking the factor space.  So, in
this case, there are two equivalent, `dual' perspectives of the
reduction process, the one using presymplectic geometry in an
intermediary step and the other one using Poisson geometry.

\smallskip

The case of a general Hamiltonian system with a closed constraint
algebra is now basically a combination of the previous two
diagrams. The essential point to prove is again under which
circumstances the reduced phase space $(R,\o_R)$ is isomorphic to (or
at least a covering of) an appropriate orbit space.

\begin{theo}
\label{main} Let $\Phi^{\alpha}$, $1\leq\alpha\leq d$ be regular
constraint functions with nonempty constraint surface
$C=\phi^{-1}(0)$ on a connected symplectic manifold $(M,\o)$ with a
complete Poisson map $\phi\colon (M,\o)\to
(P,\Pi_P),x\mapsto\Phi^{\alpha}(x)$, and denote the symplectic leaf
through $0\in P$ by $L_0$ and its pre-image under $\phi$ in $M$ by
$M_0:=\phi^{-1}(L_0)$. The orbit space of $M$ ($M_0$) with respect to
the flow generated by the constraints $\Phi^{\alpha}$ is denoted by
$Q$ ($Q_0$).

If $Q$ is a manifold, then the reduced phase space $R$ is a
(symplectic) covering of $Q_0$, which is a symplectic leaf of $Q$.  If
$\pi_1(L_0)=0$, this covering is a symplectomorphism.
\end{theo}

\begin{equation} \mbox{\parbox[c]{15cm}{
\begin{picture}(120,32)(-18,40)
\put(-19.5,70){$(C=\phi^{-1}(0),\omega_C) \hspace{1mm}
\stackrel{\iota_1}{\longrightarrow} \hspace{1mm}
(M_0=\phi^{-1}(L_0),\omega_{M_0})\
\stackrel{\iota_0}{\longrightarrow}\ (M,\omega)$}
\put(63,71.8){\rotatebox{90}{\oval(1.5,1.5)[t]}}
\put(14,71.8){\rotatebox{90}{\oval(1.5,1.5)[t]}}
\put(35,68){\vector(0,-1){20}} \put(-3,68){\vector(0,-1){20}}
\put(76,68){\vector(0,-1){20}}
\put(80,68){\vector(3,-2){28}}
\put(-17,42){$(R=C/\!\!\sim,\omega_R)\
\stackrel{\cong}{\longrightarrow} \
(Q_0=M_0/\!\!\sim,\omega_{Q_0})\!\!\hspace{3.5mm} \longrightarrow
\hspace{3.5mm} (Q=M/\!\!\sim,\Pi_Q) \hspace{6mm}(P,\Pi_P)\
\hookleftarrow\ (L_0,\omega_{L_0})$}
\put(58.5,43.9){\rotatebox{90}{\oval(1.5,1.5)[t]}}
\put(37,57){$\pi|_{M_0}$}\put(-1,57){$\pi|_{C}$}
\put(77,57){$\pi$}
\put(100,57){$\phi$}
\end{picture}}}\label{diagr:mixed}
\end{equation}

The standard reduction procedure consists in first going to the
presymplectic\footnote{Within this subsection `presymplectic' not
necessarily contains the condition that the kernel of the closed
2-form has constant rank.} constraint surface $(C,\o_C)$ embedded by
$\iota = \iota_0 \circ \iota_1$ into $M$ and then taking the factor
space with respect to the orbits generated by the vector fields in the
kernel of $\o_C$. (By a slight abuse of notation we denoted the
corresponding equivalence relation by $\sim$, too, while the flow of
all of the constraints $\Phi^\alpha$ certainly does not remain inside
of $C$). The second step is equivalent to restricting the projection
map $\pi\colon M \to Q$ to $C \subset M$. According to the theorem
(provided the covering map from $R$ to $Q_0$ is an isomorphism), we
may alternatively restrict to a generically much larger (also
presymplectic) submanifold $(M_0,\o_{M_0})$ of $M$ and then, in a
second step, factor out the flow generated by {\em all\/} the
constraints $\Phi^\alpha$. (Note that not all the vector fields
$v^\alpha$ generating this flow are in the kernel of
$\o_{M_0}$. Below, however, we will define another presymplectic form
$\bar \o_{M_0}$, generalizing $\bar \o$ of Corollary \ref{ASS}, so
that the kernel of this presymplectic form is spanned by the $v^\a$.)
Finally, as obvious also from the commutative diagram
(\ref{diagr:mixed}) above, a third alternative consists in factoring
out the flow of all the constraints in a first step resulting in
the Poisson manifold $Q$, in which $Q_0$ is embedded as a symplectic
leaf.

In the second approach the first step amounts to solving only the
first class part of the constraints. Solving the second class part of
them is traded for taking the flow of all the constraints instead of
just the part of the flow which remains inside $C$. Clearly for this
alternative to work the closure of the constraint algebra is essential
since only then the Hamiltonian vector fields of the constraints are
in involution and thus generate orbits in $M$.

\smallskip

\begin{proof} Locally, one can always split the constraints into first
class and second class ones. This corresponds to choosing coordinates
in a neighborhood of $0\in P$ which are adapted to the leaf $L_0$:
First we choose arbitrary coordinates of $L_0$ and supplement them by
additional local functions in the kernel of the Poisson tensor $\Pi$
such that, taken together, they form a coordinate system in a
neighborhood of $0$. If the transition from the original coordinates
$\Phi^{\alpha}$ to the adapted ones is nonsingular (\ie with
nonvanishing Jacobian), the new coordinates form local regular and
irreducible constraints. By choosing the adapted coordinates we have
performed a local splitting of the constraints: the coordinates of the
symplectic leaf $L_0$ are of second class whereas the remaining
coordinates are of first class.

 Restricting $M$ to $M_0$ amounts to solving the first class part of
 the constraints implying that $M_0$ is a coisotropic submanifold of
 $M$. It then follows as in the purely first class case at the
 beginning of this subsection that $Q_0$ is a symplectic leaf in $Q$.
 Note that the codimension of $M_0$ in $M$ and of $L_0$ in $P$ always
 coincide due to regularity and irreducibility of the constraints,
 which implies that the differential of the map $\phi$ is
 nonvanishing.

 To obtain the reduced phase space $R$, it remains to factor out the
 flow generated by the first class constraints and to solve the second
 class constraints (in a local splitting). The first part of this
 procedure is obviously contained in factoring out by the equivalence
 class $\sim$, and analogously to the purely second class case the rest
 of the equivalence relation serves to solve the second class
 constraints.

 As in Theorem \ref{dual}, $R$ and $Q_0$ are in general not identical,
 even if $Q_0$ is a manifold. In the presence of Gribov copies $Q_0$
 is obtained from $R$ by a discrete set of identifications such that
 $R\to Q_0$ is a covering if $Q_0$ is a manifold (which is always the
 case if there is a finite number of such identifications). Using
 Lemma \ref{bundle} one sees that $M_0$ is a fiber bundle over $L_0$
 with a natural flat connection whose holonomy group determines the
 number of Gribov copies. Hence, a sufficient condition for the
 absence of identifications is simply connectedness of $L_0$.
\end{proof}

As in Corollary \ref{ASS} we may equip $M_0$ with a presymplectic form
$\bar \o_{M_0}$ such that its kernel spans the tangent space of the
fibers of $\pi \colon M_0 \to Q$. The image of $M_0 \subset M$ under
$\phi$ is $L_0$. Subtracting from (the presymplectic form)
$\omega_{M_0}$ the pull back of $\o_{L_0}$ under this map (\ie under
$\phi \circ \iota_0$) we obtain a presymplectic 2-form $\bar
\omega_{M_0}$ such as we did in Corollary \ref{ASS}. Note that the
dimension of the kernel of $\bar \omega_{M_0}$ equals $d$, the number
of constraints, and thus in particular is constant on all of
$M_0$. (The proof follows along the same lines as in the symplectic
case.)

As in the previous cases also here there is a Poisson perspective of
the reduction process, dual to the one using either the presymplectic
manifold $(M_0, \bar \o_{M_0})$ {\em or\/} the presymplectic manifold
$(C,\o_C)$.

In the particular case of $P$ being symplectic, $\iota_0$ becomes an
isomorphism, $M_0 \cong M$, and likewise $Q_0 \cong Q$, so that the
middle part of the diagram (\ref{diagr:mixed}) may be dropped
(disappears). Furthermore, the presymplectic form $\o_C$ and the
Poisson bivector $\Pi_Q$ become nondegenerate, rendering both spaces
symplectic, $\pi_C$ becomes an isomorphism, and the diagram
(\ref{diagr:mixed}) reduces to the one of the previous subsection,
diagram (\ref{diagr:second}).

Let us finally note that the problem of Gribov copies (in the sense
used in this paper) can occur only if the constrained system is not
purely of first class. For first class constraints, the equivalence
relation $\sim$ is exactly what is needed for the standard constraint
reduction; only when there are also second class constraints can a
discrete set of surplus identifications occur. The description of
systems with constraints of both first and second class (or more
generally of mixed type) described in the preceding paragraphs
suggests to take those identifications seriously and to identify the
physical phase space with $Q_0=M_0/\!\!\sim$ in all cases. The
additional identifications can then be interpreted as large gauge
transformations which appear together with the gauge transformations
generated by first class constraints.

\subsection{Transforming second class constraints to a first class
system}
\label{subsecAppl}

In the present subsection we extend on the considerations at the end
of Sec.~\ref{sec:Second} with the aim to reformulate the constrained
system with symplectic $P$ as a first class system in an extended
phase space $\widetilde M$. The setting is that of Corollary 2' with a
simply connected $P$, the original phase space $M$ being a trivial
fiber bundle with base space $R\cong C$ and fiber $P$, endowed with a
canonical flat connection (Lemma \ref{bundle}).  Moreover, with the
respective two projections, $(C,\omega_C)$ and $(P,\O_P)$ form a
symplectic dual pair with respect to $(M,\omega)$ (Theorem \ref{dual}
and Corollary \ref{simplyconn}).  The simply connectedness is a
sufficient condition for the orbit space of $(M,\bar \o)$ to coincide
with the reduced phase space $(C,\omega_C)$; otherwise the former can
be also a covering of the latter, giving rise to `Gribov copies'. (Cf
also diagram (\ref{diagr:second}). The considerations below apply also
to the slightly more general case where $\pi_1(P)\not=0$ but still the
holonomy of the flat connection is trivial such that $Q\cong R$.)  If
the assumptions of Corollary 2' are fulfilled, there are no Gribov
problems and the reduced phase space $(C,\omega_C)$ can be identified
with the orbit space of the presymplectic manifold $(M,\bar{\omega})$
\cite{ASS}. This orbit space, in turn, is obtained by symplectic
reduction (using first class constraints) of an {\em extended\/} phase
space which can be constructed in, e.g., one of the following two
ways.

\subsubsection{$(M,\bar \o)$ as graph of the constraint map}
\label{graph}

Let $(\widetilde{M}_1,\tom_1):=(M,\o)\times (P,-\O_P)$ and embed $M$
in $\widetilde{M}_1$ as the graph $\Gamma_{\phi}$ of the map
$\phi\colon M\to P$, \ie $M\to \{(x,\phi(x)):x\in M\}\subset
\widetilde{M}_1$.\footnote{We are grateful to P.~Bressler for
suggesting this construction.} As a direct product of two symplectic
manifolds $\widetilde{M}_1$ is symplectic with the product symplectic
structure. Note that we take $P$ with the symplectic structure
reversed, \ie $\tom_1=\pi_M^*\o-\pi_P^*\O_P$ where the projections
from $\widetilde{M}_1$ to $M$ and $P$, respectively, are denoted by
$\pi_M$ and $\pi_P$. This ensures that the embedding of $M$ into
$\widetilde{M}_1$ as the graph of $\phi$ is coisotropic owing to
Lemma~\ref{PoissonGraph}, which can also be checked directly: As a
subset, $M$ is specified by the constraints
$\Psi^{\alpha}(x,\Phi):=\Phi^{\alpha}-\Phi^{\alpha}(x)=0$ where
$\Phi^{\alpha}$ are coordinates of $P$ and $\Phi^{\alpha}(x)$ is the
respective original constraint as a function on $M$. The new
constraints satisfy
\begin{equation}\label{Psibracket}
\{\Psi^{\alpha}(x,\Phi),\Psi^{\beta}(x,\Phi)\}_{\widetilde{M}_1}=
-\{\Phi^{\alpha},\Phi^{\beta}\}_P+
\{\Phi^{\alpha}(x),\Phi^{\beta}(x)\}_M=
-\Pi_P^{\alpha\beta}(\Phi)+\Pi_P^{\alpha\beta}(\phi(x))\approx 0
\label{ext1} \ee by definition of $\Pi_P$ (``$\approx0$'' denotes
vanishing on the constraint surface). This verifies that $M$ is
coisotropically embedded as the graph of $\phi$ in $\widetilde{M}_1$,
and therefore the second class constraints $\Phi^{\alpha}$ can be
replaced by first class constraints $\Psi^{\alpha}$. The constraint
surface $\Psi^{\alpha}=0$ is the presymplectic manifold $(M,\bar{\o})$
(rather than the symplectic manifold $(M,\o)$) and factoring out the
kernel of $\bar{\o}$ leads to the reduced phase space $(R,\omega_R)$
according to Corollary 2'.

We thus obtain
\begin{prop} \label{prop:M1}
 The reduced phase space $(R,\o_R)$ of the second class constrained
 system $(M,\o)$ with constraint map $\phi\colon M\to P$, $P$
 symplectic and simply connected, is symplectomorphic to the reduced
 phase space of the first class constrained system
 $(\widetilde{M}_1,\tom_1):=(M,\o)\times (P,-\O_P)$ with the graph of
 $\phi$ as constraint surface.  
\end{prop}

Note that the Poisson bracket (\ref{Psibracket}) of two first class
constraints $\Psi^{\alpha}$ in general vanishes only on the constraint
surface $\Phi=\phi(x)$. Correspondingly, in the extended phase space
$\widetilde M_1$ the new constraints are first class, but, in general,
they do no longer form a closed Poisson subalgebra. In fact, they form a
closed algebra, {\em iff\/} $\Pi_P$ is at most linear in $\Phi$ (while
still respecting $\det \Pi_P^{\a\b} \neq 0$, if necessary with a
restricted $M$), \ie in the case of some centrally extended Lie
algebras.  In contrast, the second construction introduced now,
despite being more complicated at first sight, always leads to {\em
abelian\/} (\ie Poisson commuting) first class constraints.

\subsubsection{Extension by a Whitney sum}

According to Lemma \ref{bundle}, $M$ is a fiber bundle over $P$. So
are $P$ itself (with trivial fiber) and $T^*P$, and we can form the
Whitney sum of these bundles to obtain a new fiber bundle
$\widetilde{M}_2:=M\oplus P\oplus T^*P$ with base $P$ and fiber $C\times
\R^d$:

\begin{defi} Let $E_1\stackrel{\pi_1}{\longrightarrow} B$ and
$E_2\stackrel{\pi_2}{\longrightarrow} B$ be two fiber bundles over the
same base manifold $B$. The {\em Whitney sum\/} $E_1\oplus E_2$ is
defined as the fiber bundle over $B$ given by
\[ E_1\oplus E_2=\{(u_1,u_2)\in E_1\times
E_2:\pi_1\times\pi_2(u_1,u_2)=(p,p)\}\,.
\]
\end{defi}

If the bundles $E_1$ and $E_2$ are equipped with presymplectic forms
$\omega_1$ and $\omega_2$, respectively, we can also define
$(E_1,\omega_1)\oplus (E_2,\omega_2)$ as a sum of presymplectic
manifolds which will be equipped with a new presymplectic form:

\begin{defi} Let $(E_1,\omega_1)$ and $(E_2,\omega_2)$ be
presymplectic manifolds which are simultaneously fiber bundles over
the same base manifold. Then the sum $(E_1,\omega_1)\oplus
(E_2,\omega_2)$ is the fiber bundle $(E_1\oplus
E_2,\omega_1\oplus\omega_2)$ with
\[ \omega_1\oplus\omega_2:=p_1^*\omega_1+p_2^*\omega_2
\] where $p_i\colon E_1\oplus E_2\to E_i$ are the projections to the
respective factors.
\end{defi}

In our case, we have the three symplectic manifolds $(M,\omega)$,
$(P,\Omega_P)$ and $T^*P$ which all are fiber bundles over $P$. We
define the extended phase space
$(\widetilde{M}_2,\widetilde{\omega}_2):= (M,\omega)\oplus
(P,-\Omega_P)\oplus T^*P$. Note that $P$ enters with negative
symplectic structure, which implies that the presymplectic form on the
sum of the first two spaces is just $\bar{\omega}$ [note that $M\oplus
P=M$ topologically, and the projections are $(M\oplus P\to M)={\rm
id}$, $(M\oplus P\to P)=\phi$].  In addition to the projection
$p_1\colon\widetilde{M}_2\to M$ we have a canonical embedding $i\colon
M\to \widetilde{M}_2$ as the zero section in the $T^*P$-part of the
fiber.  On the extended phase space we also add the symplectic form of
$T^*P$ which results in a {\em symplectic\/} form
$\widetilde{\omega}_2$:

\begin{lemma} 
 The form $\widetilde{\omega}_2=p_1^*\bar\o+\md\Phi^{\alpha} \wedge
 \md\pi_{\alpha}$ on $\widetilde{M}_2=M\oplus\overline{P} \oplus T^*P$
 is nondegenerate ($\md\Phi^{\alpha} \wedge \md\pi_{\alpha}$ is the
 symplectic form on $T^*P$ pulled back to $\widetilde{M}_2$ via
 $p_3$).
\end{lemma}

\begin{proof} In this proof we will denote any Hamiltonian vector
field generated by a function $f$ using the symplectic structure
$\omega$ on $M$ by $X_f$. It can be transported to vector fields on
$i(M)\subset\widetilde{M}_2$ by push forward with the canonical
embedding $i$. Choosing local coordinates $y^i$ on the fibers of
$\phi$ and $\Phi^{\alpha}$ in $P$, we then have $X_{\Phi^{\alpha}}
\lrcorner \omega= \md\Phi^{\alpha}$, $X_{y^i}\lrcorner \omega=\md y^i$
and, by definition of $\Omega_P$, $X_{\Phi^{\alpha}}\lrcorner
\md\Phi^{\beta}= \{\Phi^{\beta},\Phi^{\alpha}\}=
(\Omega_P^{-1})^{\alpha\beta}$. As already seen in Corollary \ref{ASS},
$X_{\Phi^{\alpha}}\lrcorner\bar{\omega}=0$ and thus
\begin{equation}\label{Phiin}
 i_*X_{\Phi^{\alpha}}\lrcorner\widetilde{\omega}_2=
 i_*X_{\Phi^{\alpha}} \lrcorner(p_1^*\bar\o+ \md\Phi^{\beta}
 \wedge\md\pi_{\beta})= (\Omega_P^{-1})^{\alpha\beta}
 \md\pi_{\beta}\qquad \mbox{on $i(M)$}\,. \label{tildefluss}
\end{equation} 
Since $\partial/\partial y^i \lrcorner
(\widetilde{\omega}_2-\omega)=0$,
\begin{equation}\label{xin} i_*X_{y^i}\lrcorner\widetilde{\omega}_2= \md
y^i+ m_{i\alpha}\md\Phi^{\alpha}+
n_i^{\alpha}\md\pi_{\alpha}\qquad \mbox{on $i(M)$}
\end{equation} with some coefficient functions $m_{i\alpha}$ and
$n_i^{\alpha}$.  Finally, on all of $\widetilde{M}$ (not just on $i(M)$)
\begin{equation}\label{piin}
\frac{\partial}{\partial\pi_{\alpha}}\lrcorner \, \widetilde{\omega}_2=
-\md\Phi^{\alpha}
\end{equation} which together with (\ref{Phiin}) and (\ref{xin})
demonstrates that $\widetilde{\omega}_2^{\#}\colon T\widetilde{M}\to
T^*\widetilde{M}$ is surjective and hence (due to equal dimension of
domain of definition and image) bijective on $i(M)$. By continuity of
$\tom_2$, this holds true in a neighborhood of $i(M)$, and translation
invariance of $\tom_2$ in the $\pi_{\alpha}$-directions (according to
(\ref{piin}), $\partial/\partial\pi_{\alpha}$ are Hamiltonian vector
fields) implies that $\tom_2$ is nondegenerate on all of
$\widetilde{M}_2$.
\end{proof}

Important in physical applications is also the following observation:

\begin{lemma}
\label{pilemma}
Let $\pi_{\alpha}$, $1\leq\alpha\leq d$ denote the pull back by
$p_3\colon \widetilde{M}_2\to T^*P$ of the momenta on $T^*P$. Then
$\{\pi_{\alpha},\pi_{\beta}\}=0$ on
$(\widetilde{M}_2,\widetilde{\omega}_2)$.
\end{lemma}

\begin{proof} According to (\ref{Phiin}), on the constraint surface
$i(M)$ the variable $\pi_{\alpha}$ generates the Hamiltonian vector
field $X_{\pi_{\alpha}}= \left(\Omega_P\right)_{\alpha\beta}
i_*X_{\Phi^{\beta}}$ (where $X_{\Phi^{\alpha}}$ is the vector field
generated by $\Phi^{\alpha}$ using $\omega$ on $M$) using the
symplectic structure $\widetilde{\omega}_2$. Therefore,
$\{\pi_{\alpha},\pi_{\beta}\}|_{i(M)}=
i_*X_{\pi_{\beta}}\pi_{\alpha}=0$.  Since
$\partial/\partial\pi_{\alpha}$ are Hamiltonian vector fields on
$\widetilde{M}_2$, we have ${\cal L}_{\partial/\partial\pi_{\gamma}}
\{\pi_{\alpha},\pi_{\beta}\}= \{{\cal
L}_{\partial/\partial\pi_{\gamma}} \pi_{\alpha},\pi_{\beta}\}+
\{\pi_{\alpha},{\cal L}_{\partial/\partial\pi_{\gamma}}
\pi_{\beta}\}=0$, which implies $\{\pi_{\alpha},\pi_{\beta}\}=0$ on
all of $\widetilde{M}_2$.
\end{proof}

Collecting the results we obtain

\begin{prop}
\label{prop:M2} The reduced phase space $(R,\o_R)$ of the second class
constrained system $(M,\o)$ with constraint map $\phi\colon M\to P$,
$P$ symplectic and simply connected, is symplectomorphic to the
reduced phase space of the constrained system $(\widetilde{M}_2,\tom_2):=
(M,\omega)\oplus \bar{P}\oplus T^*P$ with the momenta of $T^*P$ as
abelian first class constraints.
\end{prop}

\subsubsection{Possible application and generalization}

On $\widetilde{M}_2$ we can use local coordinates $(x^i,\pi_\alpha)$
in which the symplectic form is $\widetilde{\omega}_2(x,\pi) = \bar
\omega(x) + \md \Phi^\a(x) \wedge \md\pi_\alpha$ and the constraints
are just $\pi_\alpha = 0$. (Note that by construction
$\widetilde{M}_2$ is $M \times \R^d$ globally if and only if
$T^*P\cong P\times\R^d$, i.e., if $P$ is parallelizable.) Despite of
the simple appearance of $\pi_\alpha$ in the symplectic form, it is
quite nontrivial that the constraints Poisson commute (on all of
$\widetilde{M}_2$). In comparison, the first extension
$\widetilde{M}_1$ is always a product manifold $\widetilde{M}_1 \cong
M \times P$, albeit with a topologically more complicated factor ($P$
may have nontrivial topology, even if embedded into $\R^d$).  This
extension, however, has nonabelian first class constraints, in
general they may even have nontrivial structure {\em functions}.
Various quantization schemes, such as BRST quantization (cf., e.g.,
\cite{Henneaux}), simplify greatly in the case of Poisson commuting
constraints.

In both approaches the original second class system is reformulated as
a first class system. The original constraint surface $C \cong R$ is
{\em one\/} admissible gauge fixing surface in the new
system. Choosing this gauge, \ie $\Phi^\a(x)=0$, in a quantization
scheme relying on a gauge fixation, such as some path integral
quantization schemes (Faddeev--Popov procedure and its BRST- or
BV-generalization), one reobtains the path integral formulation of the
original second class system (as one may verify {\em for both
approaches\/} by straightforward explicit calculations).  However, now
one has the option to choose {\em another\/} gauge, which may greatly
simplify the resulting path integral measure in concrete applications.
Or one can use a quantization scheme that does not need any gauge
fixing, such as Dirac quantization \cite{Dirac} (\cf also
\cite{Henneaux}), where one regards physical states as those which are
in the kernel of the quantum operators corresponding to
the constraint functions; this scheme is applicable only for a first
class constraint system (the first class property remaining alive even
after quantization, i.e.\ assuming the absence of anomalies). 

The reformulation in Proposition \ref{prop:M2} may be generalized by
relaxing the constraints $\pi_\alpha \approx 0$ to, e.g., $\pi_\alpha
-f_\a(x) \approx 0$ for some choice of $f\colon M
\to P$ such that the new system of constraints is still first class in
$(\widetilde{M}_2, \widetilde{\o}_2)$. It is possible that such more
complicated constraints lead to a simplification in the final result
for a path integral measure after fixing the gauge.

It does not seem straightforward (although interesting) to us how to
extend the first class reformulation of the present subsection to the
general case of a Poisson manifold $P$ (not necessarily symplectic).
If, on the other hand, the constraints $\Phi^\alpha$ may be split
globally (on all of $M$, not just in a neighborhood of $C$) into first
class and second class constraints, then essentially in the above
extension one merely has to replace $(P,\O_P)$ by the symplectic leaf
$(L_0,\o_{L_0})$ and the extension goes through without difficulties.

In order to transform a complete physical system from second class to
first class, one also has to find observables (in particular this
applies to the Hamiltonian) which are in involution with the new
constraints. In general, this can be done as follows: Observables of
second class systems are arbitrary functions on $M$, where, however,
only the restriction to the constraint surface is of physical
significance. This means that independent observables correspond to
arbitrary functions on $Q$. Using the projection $\pi$ they can be
pulled back to $M$ and then extended in an arbitrary way to the
extended phase space to yield a set of functions there. By
construction, these functions are in involution with the new
constraints on the `constraint surface' $M$ (they are constant on the
orbits of $M$) and so are observables of the first class system. See
also \cite{Marnelius} for a discussion of observables in a nonstandard
approach of handling second class constrained systems.

The two proposals for turning a second class constrained system into a
first class one may be compared to other methods in the
literature. First, there is the Faddeev--Shatashvili approach
\cite{FS} of handling second class constraints. In fact, for their toy
model $(M\cong T^*(\R^2), \md x \wedge \md p_x + \md y \wedge \md
p_y)$ constrained by $\Phi^1 = -x$, $\Phi^2 = -p_x$, Prop.\
\ref{prop:M1} specializes to \cite{FS}. However, for their more
complicated, realistic system of physical interest, the constructions
differ on several grounds: There are, e.g., topological differences
between the extended phase spaces; but above all, their original
system of second class constraints does not form a closed Poisson
subalgebra, and thus the constructions developed in the present paper
(and also in \cite{ASS}) do not even apply.

Another method for turning a second class constrained system into a
first class one has been introduced in \cite{Batalin}. Besides working
only locally, the new first class constraints are given there by an
iteration procedure which does not allow to find the constraints in
closed form for a general system. An alternative has been mentioned in
\cite{Batalin2} where the extended phase space is simply
$\widetilde{M}=M\times\R^d$ with symplectic form
$\widetilde{\omega}=p_1^*\omega_M+(p_2^*\md \pi_{\alpha})\wedge
(p_1^*\md \Phi^{\alpha})$ (in our notation, denoting the projections
from $\widetilde{M}$ to $M$ and $\R^d$ by $p_1$ and $p_2$,
respectively; $\pi_{\alpha}$ are global coordinates of $\R^d$). This
procedure obviously works globally, and it also turns the second class
constraints $\Phi^{\alpha}$ into abelian first class constraints
$p_1^*\Phi^{\alpha}$ on $\widetilde{M}$ (the fact that the constraints
on $\widetilde{M}$ are abelian follows easily from $v^{\alpha}=
X_{p_1^*\Phi^{\alpha}}= \frac{\partial}{\partial
(p_2^*\pi_{\alpha})}$). However, here we prefer the extension
$\widetilde{M}_2$, which also leads to an abelian first class system,
because according to (\ref{piin}) it preserves the orbits generated by
the original second class constraints on $M$, whereas the method of
\cite{Batalin2} leads to orbits $\R^d$ in the constraint surface which
are always of trivial topology. (Note that in this extension it is not
possible to choose $p_2^*\pi_{\alpha}$ as first class constraints with
$\Phi^{\alpha}=0$ as gauge fixing conditions since the constraint
manifold $p_2^*\pi_{\alpha}=0$ would be $(M,\omega)$ and so
nondegenerate.)

\section{Dirac brackets and leaf-symplectic embeddings of Poisson
manifolds}
\label{sec:LeafSymp}

In the previous section we have seen the appearance of Poisson
manifolds in the context of a constrained system with closed
constraint algebra. In the pure second class case, however, both $P$
and $Q$ were {\em symplectic}. On the other hand, we noted that the
original manifold $M$ can be equipped naturally with a {\em
presymplectic\/} form (\ref{baro}) such that, under favorable
circumstances specified above which guaranteed the absence of a
`Gribov problem', the constraint surface is an admissible cross
section of the respective orbits.

There is, however, another perspective, in a way {\em dual\/} to
the one with a presymplectic manifold $(M, \bar \omega)$ and which is
applicable for any second class constrained system (not necessarily
forming a closed algebra), where a Poisson manifold plays a role.
This is the concept of the well-known Dirac bracket
\cite{Dirac}.

\subsection{Dirac bracket}
\label{sec:DirBr}

Given a symplectic manifold $(M,\omega)$ with a system of (regular,
irreducible) second class constraints $\Phi^\a$, $\a=1,\ldots,d$, $\{
\Phi^\a , \Phi^\b \} (x) = F^{\a\b}(x)$, $\det F^{\a\b} \not \approx
0$, Dirac defined the so called Dirac bracket as a modified Poisson
bracket $\{ \cdot , \cdot \}_D$ on $M$. We again denote the constraint
surface $\Phi^\a(x)=0$ by $C$ and the induced {\em symplectic\/}
2-form by $\o_C$; if $\iota_C\colon C \emb M$ is the respective
embedding map, then $\o_C = \iota_C^* \o$. Using the Poisson bivector
$\Pi$ inverse to $\omega$, the bivector $\Pi_D$ corresponding to the
bracket $\{ \cdot , \cdot \}_D$ has the form \be \Pi_D = \Pi + \2
G_{\a\b} v^\a \wedge v^\b \; , \label{pd} \ee where $G_{\a\b}$ is the
inverse to $F^{\a\b}$, well-defined at least in some tubular
neighborhood $S\subset M$ of the constraint surface, $C \subset S$,
and $v^\a = \{ \cdot , \Phi^\a \}\equiv - (\partial_i \Phi^\a)
\Pi^{ij} \partial_j$ is the Hamiltonian vector field generated by the
constraint $\Phi^\a$.

One may verify that $\Pi_D$ indeed satisfies the Jacobi identity and
thus defines a Poisson bracket.\footnote{This may be infered also from
the considerations to follow. With Prop.~\ref{prop:Dirac} one finds
that the leaves $\Phi^\alpha = {\mbox{const}}$ of the local foliation
generated by $\Pi_D$ are symplectic which shows that $\Pi_D$ is
Poisson: In a neighborhood $S$ of $C$, $\Pi_D$ has constant rank and
is surface forming. This implies \cite{inprep} that there is a 3-form
$H$ on $S$ such that $[\Pi_D,\Pi_D]=(\Pi_D^{\sharp})^{\otimes 3}H$
(such a bivector $\Pi_D$ is called $H$-Poisson, cf.~also the remark
after Def.~\ref{compat}). On the leaves of $\Pi_D$, the exterior
derivative of its inverse equals the pull back of $H$. Using our later
result (Prop.~\ref{prop:Dirac}) this implies that the pull back of $H$
to an arbitrary leaf in $S$ vanishes, and so $H$ is purely transversal
with respect to the foliation. The transversal part, however, is
projected out by $(\Pi_D^{\sharp})^{\otimes 3}$ which gives us
$[\Pi_D,\Pi_D]=0$.}  By construction the constraint functions
restricted to $S$, $\Phi^\a|_S \in C^\infty(S)$, span the center of
the algebra generated by this bracket. Correspondingly the constraint
surface is a symplectic leaf of $\Pi_D$. Moreover, the symplectic form
on that leaf coincides with $\o_C$ (we will verify this explicitly in
Proposition \ref{prop:Dirac} below.)

Inspired by the relation of the Dirac bracket of a second class
constrained system with the original symplectic structure, we make the
following definition.

\begin{defi} Let $(P,\Pi)$ be a Poisson manifold and $\Omega_L$ denote
the induced symplectic form on a given leaf $L$ of the stratification
of $P$ with $\iota_L\colon L \emb P$ being the respective embedding
map. A (symplectic, presymplectic, $\ldots$) 2-form $\widetilde \Omega$
and a Poisson bivector $\Pi$ are called {\em compatible within\/} $S
\subset P$, if for any leaf $L$ in $S$: $\O_L = \iota_L^* \widetilde \O$.
\label{compat}
\end{defi}

\begin{rem} There is an obvious adaptation of the above definition to
the case of an almost Poisson bivector $\Pi$ which still generates a
foliation (stratification)---so called $H$-Poisson or twisted Poisson
structures $\Pi$, fulfilling
$[\Pi,\Pi]=\left(\Pi^\sharp\right)^{\otimes 3} \! H$ for some
closed 3-form $H$ \cite{almost,Park,SeWe,inprep}. 
\end{rem}

The Dirac bivector (\ref{pd}) is well-defined in some subset $S
\subset M$ containing the constraint surface $C$. Within $S$ we have
the following relation

\begin{prop}
\label{prop:Dirac} The Dirac bivector $\Pi_D$ and the symplectic form
$\o \in \Lambda^2(T^*M)$ are compatible within the domain $S \subset
M$ of definition of $\Pi_D$.
\end{prop}

\begin{proof} It suffices to show that for any two functions $f$ and
$g$ on a given leaf $L$ of the foliation with embedding map $\iota_L :
L \to S$ one has on all of $L$
\be \iota_L^* \left(\{F,G\}_D\right) = \widetilde X_g (f) \; ,
\label{zwischen} \ee where $F$ and $G$ are arbitrary extensions of $f$
and $g$ to $S$, such that $\iota_L^* F = f$ and $\iota_L^* G = g$, and
where $\widetilde X_g \in \Gamma(TL)$ is (uniquely) defined by means
of $\widetilde X_g \ins \iota_L^* \, \omega = \md g$.
(The right hand side is the Poisson bracket between $f$ and $g$ as
induced by $\iota_L^* \omega$, while on the other hand the left-hand
side gives the corresponding bracket induced by the symplectic form of
the bivector $\Pi_D$ on $L$. Thus, equality for all functions $f$ and
$g$ proves equality of the respective 2-forms.)

We first note that the choice of the extension of the function $f$
(and likewise of $g$) does not enter the left-hand side of Eq.\
(\ref{zwischen}), since by construction $\{ \cdot , \Phi^\a \}_D
\equiv 0$, while two different extensions $F$, $F'$ of $f$ differ only
by a combination of the constraints: $F' = F + c^\a \Phi_\a$ for some
smooth functions $c^\a$ defined in a neighborhood of $L$ (cf.\ e.g.\
\cite{Henneaux}).  We thus may choose a particularly convenient
extension, namely one such that $\iota^*_L \left(v^\alpha(F)\right)
=0$ (and likewise for $G$), where, as before $v^\alpha = \{ \cdot ,
\Phi^\a \}$. It then immediately follows from the definition
(\ref{pd}) of the Dirac bracket that $\iota_L^*\{F,G\}_D =
\iota_L^*\{F,G\} \equiv \iota_L^* \left( X_G(F) \right) \equiv X_G|_L
(f)$ where $X_G = \{ \cdot , G \}$, since, due to the chosen
extension, $X_G$ is tangential to $L$ and thus may be restricted
consistently to $L$.  {}From $X_G|_L\ins\iota_L^*\o=\iota_L^*\left(X_G
\ins \omega \right)= \iota_L^*\md G \equiv \md g$ it thus follows
that $X_G|_L = \widetilde X_g$, which
proves the assertion.
\end{proof}

\noindent {\it Remarks:}

 i) The constraint surface is the preimage of the origin of the
 constraint map $\Phi^\a\colon M \to \R^d$. (Note that we do here no
 longer require that the constraint algebra is closed; thus, the target
 does not inherit canonically a Poisson bracket in this case). This
 corresponds only to one particular symplectic leaf $L_0$ of
 $(S,\Pi_D)$. If one shifts the constraint surface slightly by setting
 $\Phi^\a(x)$ to some constant $c^\a \in \R^d$ small enough such that
 the respective preimage $L_c$ is still in $S$, the bivector $\Pi_D$
 given in (\ref{pd}) still provides the respective Dirac bracket. So
 all the leaves $L_c$ of $(S,\Pi_D)$ are seen to be possible
 constraint surfaces.

 ii) Constraints describing a fixed constraint surface $L_0$ are
 defined only up to redefinitions $\Phi^\a \to \wt \Phi^\a \equiv
 A^\a{}_\b \Phi^\b$ of the constraint functions, where the coefficient
 matrix $A^\a{}_\b(x)$, required to have a nonvanishing determinant on
 $C$, is in general a smooth function on phase space (or at least on
 some neighborhood of $C$).  The corresponding Dirac bivector $\wt
 \Pi_D$, defined in some region $\wt S \subset M$ containing $C$, now
 has in general different symplectic leaves on the intersection of $S$
 and $\wt S$ to the ones generated by $(\Pi_D,S)$; in general only the
 constraint surface $C$ itself is a joint symplectic leaf.  So, one
 obtains different possible constraint surfaces $\wt L_c$, shifted as
 in the previous remark, when one redefines the constraint functions
 which specify the original constraint surface $C=L_0=\wt L_0$.  Any
 Dirac bivector $\Pi_D$ in Eq.\ (\ref{pd}), corresponding to some
 specific choice of the constraint functions $\Phi^\a$ and defined in
 a region $S \subset M$, is compatible with the original symplectic
 2-form $\omega$ within $S$.

 iii) According to the above consideration, $S$ is foliated into
 leaves $C$ (with a slight abuse of notation, since previously $C$
 denoted only the original constraint surface) which are all second
 class submanifolds, and at any point $x \in S\subset M$ we have a
 splitting of the tangent space according to $T_xM = T_x C \oplus T_x
 C^\perp$, and likewise of the cotangent bundle $T^*_xM = \Ann_M(T_x
 C^\perp ) \oplus \Ann_M(T_x C)$, induced by the original bivector
 $\Pi$ (cf.~also (\ref{TCnormal}) with $\omega^\sharp =
 -\left(\Pi^\sharp\right)^{-1}$ and recall the discussion at the end
 of the proof of Theorem \ref{dual}). Denote by $\pi_1$ the projection
 to the first factor in $T_x M$ and by $\bar \pi_1$ to the first
 factor in $T^*_xM$ and regard both $\Pi$ and $\Pi_D$ as bivector
 fields defined on $S \subset M$. Then the definition (\ref{pd}) of
 the Dirac bivector $\Pi_D$ is equivalent \cite{MarsdenRatiu} to
 saying that $\Pi_D$ is the projection of $\Pi$ to $TC$ along
 $TC^\perp$. In formulas this becomes $\Pi_D^\sharp = \pi_1
 \circ\Pi^\sharp\circ \bar \pi_1$ or, equivalently, $\Pi_D = \Pi \circ
 (\bar \pi_1 \otimes \bar \pi_1)$ or, if one regards the bivectors as
 elements or sections of $\Lambda^2 TM$, $\Pi_D = (\pi_1 \otimes
 \pi_1) \Pi$. This follows easily using previous results: As mentioned
 before Prop.~\ref{FirstSecondSymp}, we have
 $\Ann_M(TC)=\prec\!\!\md\Phi^{\alpha}\!\!\succ$ implying
 $\pi_1\circ\Pi^{\sharp}\circ\bar\pi_1(\md\Phi^{\alpha})=0$.
 Furthermore, $\pi_1\circ\Pi^{\sharp}\circ\bar\pi_1$ restricted to the
 leaves is by definition of $\bar\pi_1$ non-degenerate and coincides
 with $\Pi^{\sharp}$. So the two properties which characterize the
 Dirac bivector $\Pi_D$ in (\ref{pd}) are fulfilled which proves
 $\pi_1\circ\Pi^{\sharp}\circ\bar\pi_1=\Pi_D^{\sharp}$. This presents
 an alternative proof for the compatibility of $\Pi_D$ with the
 original symlectic form $\omega$.

\subsection{Closed constraint algebra revisited}

Before proceeding let us specialize the above considerations to a
system of a closed, second class constraint algebra. In this case the
Dirac bivector takes the form \be \Pi_D = \Pi - \2 \left(\O_P\right)_{\a\b}
(\md \Phi^\a \ins \Pi) \wedge (\md \Phi^\b \ins \Pi) \; , \label{pd2}
\ee wherever $\O_P^{\sharp} = -(\Pi_P^{\sharp})^{-1}$ is
defined, \ie at least in a tubular neighborhood $S$ of the constraint
surface $C$. For simplicity we assume $S = M$, or, equivalently,
$(P,\Pi_P)$ symplectic. (We refer to diagram (\ref{diagr:second}) for
the notation.) Then $(M,\Pi_D)$ is a Poisson manifold.

First of all we note that the expression (\ref{pd2}) obviously is 
invariant 
with respect to a (regular) coordinate change in $P$, $\Phi^\a \to
\wt \Phi^\a (\Phi)$, corresponding to a particular change of
constraint functions, \ie to a particular matrix $A^\a{}_\b(x)$ in
Remark ii) above.  By construction, the reduced phase space
$(R=\phi^{-1}(0), \o_R)$ is a symplectic leaf of this Poisson
manifold, furthermore. This also applies to any other, shifted value
for the constraint functions, $\Phi^\a := const^\a$, \cf Remark i)
above. 

The bivector $\Pi_D$, defined on all of $M$, is not only compatible
with $\o$ (\cf Proposition~\ref{prop:Dirac} above), but clearly also
with $\bar \o$ defined in (\ref{baro}). As remarked
already at the beginning of this section, $M$ equipped with the
presymplectic form $\bar \o$ provides a perspective `dual' to the one
following from $M$ being equipped with the Dirac bivector $\Pi_D$.

There is now also {\em another\/} (nonsymplectic) Poisson manifold we
can associate to the system at hand: In Sec.~\ref{subsecAppl}, and
under the assumptions stated there, we considered two reformulations
of the original second class constrained system in an extended phase
space. In the second of those extensions, $(\widetilde M_2, \widetilde
\o_2)$, the first class constraint algebra in the extended phase space
was abelian and thus in particular closed. Therefore we can adapt
diagram (\ref{diagr:first}) to the present situation (combining it
with part of diagram (\ref{diagr:second})):
\begin{equation} \mbox{\parbox[c]{7cm}{
\begin{picture}(70,39)(40,34)
\put(40,70){$(M,\bar \o)\hspace{3mm}
\stackrel{i}{\longrightarrow}\hspace{3mm} (\widetilde M_2, \widetilde
\o_2)$}
\put(56.5,71.8){\rotatebox{90}{\oval(1.5,1.5)[t]}}
\put(45,68){\vector(0,-1){20}}
\put(75,68){\vector(0,-1){20}}
\put(79,68){\vector(1,-1){20}}
\put(14,42){$(R,\omega_R)\hspace{2mm}
\stackrel{\cong}{\longrightarrow} \hspace{2mm} 
(Q,\Omega_Q)\hspace{3mm}\hookrightarrow \hspace{3mm}
(\widetilde Q,\widetilde{\Pi}_Q)\hspace{17mm}(\widetilde P=\R^d,0)$}
\put(47,57){$\widetilde{\pi}|_M=\pi$}
\put(77,57){$\widetilde{\pi}$}
\put(98,57){}
\end{picture}}}\label{diagr:neu}
\end{equation}

This plot displays a `duality' between $(M,\bar \o)$, a presymplectic
manifold, and the Poisson manifold $(\widetilde
Q,\widetilde{\Pi}_Q)$. It is therefore natural to ask, what is the
relation between the two Poisson manifolds $(M,\Pi_D)$ and
$(\widetilde Q,\widetilde{\Pi}_Q)$.  Note as an aside that the second
Poisson manifold can be defined in general only by the second of the
two extensions of Sec.~\ref{subsecAppl}. (As remarked at the end of
Sec.~\ref{graph}, the first class constraints in $\widetilde M_1$
form a closed algebra only when the original constraint algebra
$\Pi_P$ is at most linear in the constraints $\Phi^\a$.)  In this
physically particularly interesting case, however, an analogous
question may be posed.

\begin{prop} Under the conditions specified in Corollary 2', the
Poisson manifolds $(M,\Pi_D)$ and $(\widetilde
Q,\widetilde{\Pi}_Q)$ are isomorphic, iff $P \cong \R^d$
topologically.
\end{prop}

\begin{proof}
The necessity follows from the remark that $\widetilde Q \cong Q
\times \R^d$, \ie the constraints $\pi_\a$ act trivially on the local
linear factor $\R^d$ in $\widetilde{M}_2$ (with its coordinates
$\pi_\a$) and have a flow agreeing with that generated by $\md \Phi^\a
\ins \Pi$ on $M$.  This is a consequence of (\ref{tildefluss}), where
$i(M)$ may as well be taken to denote {\em any\/} section in
$\widetilde M_2$ specified by some constant value of the momenta
$\pi_\a$ (\cf also Lemma \ref{pilemma}).

The rest of the claim is proved by establishing symplectomorphisms
between the symplectic leaves $\pi_\a = const$ of $\widetilde{\Pi}_Q$
and $\Phi^\a(x)=const$ of $\Pi_D$. Using the direct product structure
of $M$ under the conditions of Corollary 2', one easily sees that the
leaves are all symplectomorphic to the reduced phase space, while the
leaf spaces are both $\R^d$.
\end{proof}

\subsection{Compatible presymplectic forms on neighborhoods of regular
 leaves}

We may now reverse the question addressed originally by Dirac: Given a
Poisson bivector $\Pi \in \L^2(TM)$ on a manifold $M$, is there a
symplectic (or, since $\dim M$ may be odd, at least presymplectic)
2-form which is compatible with $\Pi$ in $M$?

In general, such a compatible presymplectic 2-form will not exist on
all of $M$ (recall that also the Dirac bivector is defined only in a
neighborhood of the constraint surface); but it may nevertheless exist
in some neighborhood $S$ of a given leaf $L$ in $M$. In particular
$\rank \Pi$ should be constant in all of $S$, so $S$ should be a
regular Poisson manifold. For simplicity, we will assume that already
$M$ is regular and look for a compatible presymplectic form on all of
$M$.

We remark that regularity of $(M,\Pi)$ does not imply that $M$ is
foliated regularly: Consider e.g.\ the Poisson tensor $\Pi =
\left(\partial_1 + \o \partial_2 \right) \wedge \left(\partial_3 +
\bar \o \partial_4 \right)$ with irrational $\o$, $\bar \o$ on a torus
$T^4$ (coordinates $x^i \equiv x^i + 1$). Any of its symplectic leaves
is dense in all of $T^4$. This bivector is easily verified to permit the
compatible symplectic 2-form $\Omega=(1+\o\bar\o)^{-1}(\md x_1\wedge \md
x_3+\md x_2\wedge\md x_4)$ on $T^4$ (or simply $\md x_1\wedge\md x_3$
as a compatible presymplectic form).

We first construct an {\em almost\/} presymplectic form (\ie just any
2-form, not necessarily closed) which is compatible
with $\Pi$. This is possible always: Picking an auxiliary Riemannian
metric on $M$,  we can define a
2-form $\wtO \in\L^2(T^*M)$ by requiring that for any $X\in M$ a
vector normal to the leaf through $X$ is annihilated by $\wtO$
whereas for a tangential vector $v\in T_XL$ we have
$v\lrcorner\wtO=v\lrcorner\O_L$ (again, $\O_L$ is the symplectic
form induced on the leaf by $\Pi$). In this way we obtain a
well-defined 2-form on $M$ which is compatible with $\Pi$, but not
closed in general.

There is still some freedom in defining this form since one can add an
arbitrary 2-form $\lambda$ which vanishes when pulled back to a leaf.
The space of such $k$-forms is denoted by $\Lambda_0^k(M)$. This may
be exploited to obtain another compatible 2-form $\O
:=\wtO+\lambda$. Since $\md$ maps $\Lambda_0^k(M)$ to
$\Lambda_0^{k+1}(M)$ and $\md \wtO \in \Lambda_0^3(M)$, the question
of whether $\lambda$ may be chosen in such a way that $\O$ is closed,
relates to the so-called characteristic form-class of $\Pi$, \ie by
$\md \wtO$ regarded as an element of the relative cohomology $H_{\rm
rel}^3:=\Kern\md^{(3)}_0/\mbox{Im} \; \md^{(3)}_0$ ($\md_0^{(k)}$
denotes the restriction of $\md$ to $\Lambda_0^k(M)$; \cf also
\cite{VaismanBook}):

\begin{prop}
\label{prop:comp}
Let $(M,\Pi)$ be a regular Poisson manifold and $\wtO$ be any 2-form
compatible with $\Pi$ (such a 2-form exists always).

There is a compatible presymplectic form $\O$ on $M$ if and only if
the characteristic form-class of $\Pi$, $[\md\wtO]\in H_{\rm rel}^3$,
vanishes.
\end{prop}

\begin{proof}
Note first that by construction $\md\wtO\in\Lambda_0^3$ such that
$[\md\wtO]$ is well defined in the relative cohomology.

Let us first prove that the condition is necessary: assume that there
exists a compatible presymplectic form $\O$. Then $[\md\O]=0$ because
$\O$ is closed, and $\O=\wtO+\lambda$ for some
$\lambda\in\Lambda_0^2(M)$ implies
$[\md\wtO]=[\md\O]-[\md\lambda]=0$.

Conversely, if $[\md\wtO]=0$, then $\md\wtO=\md\lambda$  for some
$\lambda\in\Lambda_0^2(M)$. This implies that $\O:=\wtO-\lambda$ is
closed and compatible.
\end{proof}

In the remainder of this section we will discuss this main condition
in a more explicit form. Using the auxiliary metric, we have a
splitting of the tangent bundle $TM=TL\oplus T^\perp L$ interpreting
both bundles on the right hand side as vector bundles over the base
manifold $M$. Here, we have to assume that the normal distribution
$T^{\perp}L$ is integrable for reasons which will become clear in the
course of this subsection.

Using the metric, we obtain a dual decomposition of the cotangent
bundle $T^*M$ leading to a two-fold grading of $\Lambda^*(T^*M)$ by
taking exterior powers.  In this way the decomposition $TM=TL\oplus
T^{\perp}L$ entails a decomposition of $n$-forms
$\omega\in\Lambda^n(T^*M)$. More explicitly, we have
$\omega=\sum_{i=0}^n\omega^{(i,n-i)}$ fulfilling
$v_j\lrcorner\cdots\lrcorner v_1\lrcorner\omega_X^{(i,n-i)}=0$ for
$j>n-i$ and all $v_1,\ldots,v_j\in T_X^{\perp}L$, whereas for $j=n-i$
either $\omega^{(i,n-i)}=0$ or there are $v_1,\ldots,v_j\in
T_X^{\perp}$ such that $v_j\lrcorner\cdots\lrcorner
v_1\lrcorner\omega_X^{(i,n-i)}\not=0$.  The uniqueness of this
decomposition (for a fixed Riemannian metric) can also be seen in
local coordinates adapted to the foliation of $M$, in which it amounts
to collecting terms with a fixed number of differentials along normal
coordinates.

\begin{defi} 
 An $n$-form $\omega$ is called {\em pure of degree $(i,n-i)$\/} if it
 has the decomposition $\omega=\omega^{(i,n-i)}$. The space
 $\Lambda^{(i,n-i)}(T^*M)\subset\Lambda^n(T^*M)$ is the space of pure
 forms of degree $(i,n-i)$.
\end{defi}

Note that these and the following definitions depend on the Riemannian
metric chosen on $M$.

We can similarly decompose the exterior derivative operator $\md$ into
two parts
\[ \mdp\colon\Lambda^{(i,n-i)}(T^*M)\to\Lambda^{(i+1,n-i)}(T^*M)\mbox{
and } \mdo\colon\Lambda^{(i,n-i)}(T^*M)\to\Lambda^{(i,n+1-i)}(T^*M)\,.
\] Given a pure form $\omega\in\Lambda^{(i,n-i)}(T^*M)$, we define
\[ \md\omega=(\md\omega)^{(i+1,n-i)}+(\md\omega)^{(i,n+1-i)}=:
\mdp\omega+\mdo\omega\,.
\] Both derivative operators can be extended linearly so as to be
defined on arbitrary forms. Choosing local coordinates
$(X^{\alpha},X^I)$ such that $X^{\alpha}$ parameterize the leaves in a
neighborhood and $X^I$ the normal directions, the new derivative
operators can be written as $\mdp=\partial_{\alpha}\md
X^{\alpha}\wedge$ and $\mdo=\partial_I\md X^I\wedge$. At this point
the integrability of the normal distribution has been used: otherwise
there would be an additional term in the decomposition of $\md$ (see
also \cite{VaismanBook}). This can be seen using the Cartan formula which
for a 1-form $\omega$ reads
\[
 \md\omega(v_1,v_2)=v_1\omega(v_2)-v_2\omega(v_1)-\omega([v_1,v_2])\,.
\]
Choosing $v_1,v_2\in T^{\perp}L$ such that $[v_1,v_2]\not\in
T^{\perp}L$ (which by definition is possible only if the normal
distribution is not integrable), there is always a 1-form
$\omega\in\Lambda^{(1,0)}(T^*M)$ with
$\md\omega(v_1,v_2)\not=0$. Thus, $\md\omega$ has a nonvanishing
contribution in $\Lambda^{(0,2)}(T^*M)$ and $\md\not=\mdp+\mdo$ for a
nonintegrable normal distribution. One can see that there is only one
additional term in the general case mapping $\Lambda^{(i,j)}(T^*M)$ to
$\Lambda^{(i-1,j+2)}$; but this would already complicate the descent
equations derived below. Therefore, we will only deal with the
integrable case from now on, for which we have

\begin{lemma}\label{dsquare} 
\[
 \mdp^2=0\quad,\quad\{\mdp,\mdo\}:=\mdp\mdo+\mdo\mdp=0\quad,
 \quad\mdo^2=0
\]
\end{lemma}

\begin{proof} 
 It suffices to prove the assertion for actions on a pure form
 $\omega$ of arbitrary degree. We then have
\[
 0=\md^2\omega=\md(\mdp\omega+\mdo\omega)=\mdp^2\omega+
 \{\mdp,\mdo\}\omega+ \mdo^2\omega\,.  
\] 
 Because the three terms in the sum are all pure of different degrees,
 they have to vanish separately.
\end{proof}

Now we are in the position to proceed with the derivation of
conditions for the existence of a compatible presymplectic form. The
2-form $\wtO$ introduced above is pure of degree $(2,0)$ by
construction, $\wtO=\wtO^{(2,0)}$. As already discussed in the
paragraph preceding Proposition~\ref{prop:comp}, $\wtO$ is not
necessarily closed since $\mdo\wtO\not=0$ in general. Adding a form
$\lambda\in\Lambda_0^2(M)$ leads to a new form
$\O=\wtO^{(2,0)}+\O^{(1,1)}+\O^{(0,2)}$ which is closed if and only if
\[ 
 \md\O=\mdo\wtO^{(2,0)}+\mdp\O^{(1,1)}+\mdo\O^{(1,1)}+
 \mdp\O^{(0,2)}+\mdo\O^{(0,2)}=0\,.
\] 
Collecting forms of equal degree immediately leads to

\begin{prop}\label{Pseudo} 
 Let $(M,\Pi)$ be a regular Poisson manifold being equipped with a
 Riemannian metric and an associated integrable decomposition of the
 tangential bundle.

 There is a presymplectic 2-form
 $\O=\O^{(2,0)}+\O^{(1,1)}+\O^{(0,2)}$ compatible with $\Pi$
 on $M$ if and only if the {\em descent equations}
\begin{equation}\label{descent}
 \mdp\O^{(1,1)}=-\mdo\O^{(2,0)}\quad,
 \quad\mdp\O^{(0,2)}=-\mdo\O^{(1,1)}\quad, \quad\mdo\O^{(0,2)}=0
\end{equation} 
 have a solution on $M$ subject to the condition that $\O^{(2,0)}$
 restricted to any leaf in $M$ coincides with the symplectic form of
 that leaf.
\end{prop}

In special cases of a foliation we can reformulate the conditions of
the proposition.

\begin{cor}\label{triv} 
 If $M$ is foliated trivially, i.e.\ it is of the form $M\cong
 L\times\R^k$, then the first equation in (\ref{descent}) implies \[
 \partial_I\oint_{\sigma}\O^{(2,0)}=0 \] where $\partial_I$ denotes
 any differentiation transversal to $L$ and $\sigma$ is a closed
 two-cycle in $L$. This means that the symplectic volume of any closed
 two-cycle in a leaf has to be constant in $M$.
\end{cor}

This condition is violated for, e.g., any family of homeomorphic
coadjoint orbits of a compact, semisimple Lie algebra because the
symplectic form of leaves in the dual Lie algebra with the
Kirillov--Kostant structure depends nontrivially on the Casimir
functions (the radial coordinate, e.g., for $su(2)$). In particular,
according to the work of Kirillov, the (always discrete) set of
irreducible unitary representations of a compact, semisimple Lie
algebra corresponds to the set of all {\em integral\/} symplectic
leaves in the corresponding Lie Poisson manifold, that is to leaves
satisfying that $\oint_{\sigma}\O^{(2,0)}$ is an integer multiple of a
fixed constant for any two-cycle in the leaf.  Correspondingly,
$\oint_{\sigma}\O^{(2,0)}$ depends nontrivially on the leaf and
Poisson manifolds of this kind do not allow compatible presymplectic
forms. (In this argument we also used analyticity which implies that
$\partial_I\oint_{\sigma}\O^{(2,0)}$ cannot vanish on an interval if
$\oint_{\sigma}\O^{(2,0)}$ is not constant. This means that any open
subset of the dual Lie algebra contains (part of) a leaf violating the
condition of the Corollary.)

We have another interesting situation if all leaves in $M$ have
trivial second cohomology: $H^2(L)=0$.  If $M$ is foliated trivially,
i.e.\ of the form $M\cong L\times \R^k$, one can easily verify that
there is always a compatible presymplectic form:

\begin{lemma}\label{pot}
  If $\wtO^{(2,0)}$ has a symplectic potential $\theta^{(1,0)}$ on any
  leaf $L$ in $M\cong L\times\R^k$, i.e.\
  $\wtO^{(2,0)}=\mdp\theta^{(1,0)}$, and $\theta^{(1,0)}$ varies
  smoothly from leaf to leaf, then $\O:=\md\theta^{(1,0)}$ is a
  compatible presymplectic form on $M$.

  In particular, if all leaves $L$ in a trivially foliated $M\cong
  L\times \R^k$ have trivial second cohomology, then there exists a
  compatible presymplectic form on $M$.
\end{lemma}

This can also be derived using the descent equations which acquire the
form (using Lemma \ref{dsquare})
\[ 
 \mdp\O^{(1,1)}=\mdp\mdo\theta^{(1,0)}
\] 
solved by $\O^{(1,1)}:=\mdo\theta^{(1,0)}$, leading to
\[ 
 \mdp\O^{(0,2)}=-\mdo^2\theta^{(1,0)}=0
\] 
in addition to $\mdo\O^{(0,2)}=0$. The latter two equations have an
obvious solution $\O^{(0,2)}=0$ implying
\[
\O=\wtO^{(2,0)}+\O^{(1,1)}=\mdp\theta^{(1,0)}+ \mdo\theta^{(1,0)}=
\md \theta^{(1,0)}\,.
\]

As a more explicit example\footnote{We thank A.\ Weinstein for
suggesting to look at such an example.} we look at the manifold $M=
T^2\times \R$ with Poisson bivector $\Pi=F(x_1,x_2,x_3)
(\partial_1+\omega\partial_2) \wedge\partial_3$ with an arbitrary
function $F$ on $M$. Leaves are submanifolds subject to the condition
$\omega x_1-x_2=0$ and we can use $x_1$ and $x_3$ as local coordinates
of a leaf. If we choose the normal distribution to be spanned by
$\partial_2$, any 2-form $\Omega$ is split into $\O=
\wtO^{(2,0)}+\O^{(1,1)}+\O^{(0,2)}$ with
$\wtO^{(2,0)}=F(x_1,x_2,x_3)^{-1} \md x_1\wedge\md x_3$, $\O^{(1,1)}=
\mu_1\md x_1\wedge\md x_2+ \mu_2\md x_3\wedge\md x_2$, $\O^{(0,2)}=0$,
and only the first descent equation is nontrivial and takes the form
\[
 \mdp\O^{(1,1)}=
 (\partial_3\mu_1-\partial_1\mu_2)\md x_1\wedge\md x_2\wedge\md
 x_3=-\mdo\O^{(2,0)}= \partial_2F^{-1}\md x_1\wedge\md x_2\wedge\md
 x_3\,.
\]
This implies $\partial_3\mu_1-\partial_1\mu_2=\partial_2F^{-1}$ which
is solved by, e.g., $\mu_1=\int\partial_2F^{-1}\md x_3$, $\mu_2=0$
yielding
\[
 \O=F(x_1,x_2,x_3)^{-1}\md x_1\wedge\md x_3+\left(\int\partial_2F^{-1}\md
 x_3\right) \md x_1\wedge\md x_2
\]
as compatible presymplectic form.

Note that $\O$ is well-defined globally even if $\omega$ is irrational
in which case the leaves are dense in the torus factor of
$M$. However, if we change $M$ to be $T^2\times S^1$, the
$x_3$-integration in the second term of $\O$ is not periodic if there
is a nonvanishing zero-mode in the Fourier decomposition of
$\partial_2F^{-1}$ with respect to $x_3$. This means that in such a
case the characteristic form class of $\Pi$ on $T^2\times S^1$ does
not vanish. If $\omega$ is rational the leaves in $M$ have nontrivial
second cohomology, whereas for irrational $\omega$ the leaves are of
topology $\R\times S^1$ and so have trivial second cohomology, but $M$
is not foliated trivially. So in both cases, there is no contradiction
to Lemma \ref{pot}.

\subsection{Leafwise symplectic embeddings of Poisson manifolds}

Theorem \ref{realization} asserts that any Poisson manifold
$(P,\Pi_P)$ has a symplectic realization. This provides an appropriate
Poisson map {\em from\/} a symplectic manifold to the given Poisson
manifold. As demonstrated in Sec.~\ref{sec:Closed} above (\cf in
particular Proposition \ref{prop1} and diagram (\ref{diagr:mixed})),
physically this is, e.g., of interest in constrained systems with a
closed algebra: {\em Any\/} Poisson manifold (or at least any region
of it contained in $\R^d$, $d = \dim P$) can be understood as arising
from a constrained Hamiltonian system with appropriate constraint map
$\phi$.

The Dirac bracket and the considerations of the present section
motivate the question for another map, going in a reverse direction.
Clearly, the identity map from $(S,\o)$ to $(S,\Pi_D)$ is not a
Poisson map (here $\Pi_D$ is the Dirac bivector (\ref{pd}) defined
for some second class constraints; $S$ is the neighborhood of $C$ for
which $\Pi_D$ exists). Still, $\Pi_D$ and $\o$ are by no
means unrelated; they are what we called compatible to one another.
In the language of maps, this may be rephrased as follows: The
embedding  map from $(S,\Pi_D)$ into $(M,\o)$ is leafwise symplectic.

\begin{defi} \label{leafsymp}
 A map from a Poisson manifold $(P,\Pi_P)$ to a symplectic manifold
 $(M,\o)$ is {\em leafwise symplectic\/} (or {\em leaf-symplectic}),
 if its restriction to any symplectic leaf of $(P,\Pi_P)$ is a
 symplectic map.
\end{defi}

\begin{rem}
 Recall that according to Def.~\ref{morphium} a map $f$ between two
 symplectic manifolds $(M_1,\o_1)$ and $(M_2,\o_2)$ is symplectic, iff
 $f^* \o_2 = \o_1$. Therefore $f \colon (P,\Pi_P) \to (M,\o)$ is
 leaf-symplectic, iff $(f \circ \iota_L)^* \o = \O_L$ for any leaf $L$
 of $(P,\Pi_P)$, $\O_L$ being the induced symplectic 2-form on $L$.
\end{rem}

\begin{prop}
\label{prop:ls}
A regular Poisson manifold $(P,\Pi_P)$ permits a
leaf-symplectic embedding into some symplectic manifold, iff its
characteristic form-class vanishes.
\end{prop} 

\begin{proof}
According to Proposition \ref{prop:comp}, there exists a compatible
presymplectic form on $P$ iff the characteristic form-class of $\Pi_P$
vanishes. Assuming that there exists a leaf-symplectic embedding
$f\colon P \to M$ from $P$ to $(M,\o)$, $f^* \o$ provides a compatible
presymplectic form on $(P,\Pi_P)$. So the vanishing of the
characteristic form-class is a necessary condition. On the other hand,
if it vanishes, there exists a compatible presymplectic form $\O$ on
$P$. According to Theorem \ref{Gotay}, $(P,\O)$ can be embedded
coisotropically into a symplectic manifold $(M,\o)$. Denoting this
embedding by $\iota$ and the embedding of a symplectic leaf $(L,\O_L)$
into $P$ by $\iota_L$, we have $\O_L = \iota_L^* \O =  \iota_L^*
\iota^* \o = \left(\iota|_L\right)^* \o$. Thus the condition is also
sufficient.
\end{proof}

Leaf-symplectic embeddings can be regarded as a concept related to
isotropic symplectic realizations\footnote{We are grateful to
A.~Weinstein for making us aware of this relation.}, which are Poisson
maps $r\colon (Y,\omega)\to (M,\Pi)$ {\em from\/} a symplectic
manifold $(Y,\omega)$ to a Poisson manifold $(M,\Pi)$ such that the
fibers of $r$ are isotropic (\ie $TF\subset (TF)^{\perp}$ for any
fiber $F=r^{-1}(m)$, $m\in M$). They are of interest for symplectic
groupoids \cite{Weinstein,Karasev}, which also appeared recently in
the context of Poisson Sigma Models in Ref.\
\cite{CatFeld2}. Obstructions for the existence of an isotropic
symplectic realization have been derived \cite{DD,VaismanBook} which
are of cohomological nature and similar to those derived here for the
existence of a leaf-symplectic embedding.

While the Dirac bracket, which provided the motivation for defining
leaf-symplectic embeddings, is usually used only on symplectic
manifolds $(M,\omega)$ there is a straightforward generalization of
Def.~\ref{leafsymp} to the case of a Poisson target manifold:

\begin{defi} \label{leaftoleaf}
 A map from a Poisson manifold $(P_1,\Pi_1)$ to a Poisson manifold
 $(P_2,\Pi_2)$ is {\em leaf-to-leaf symplectic\/} if the image of any
 leaf $L_1$ of $P_1$ is symplectically embedded in a leaf $L_2$ of
 $P_2$.
\end{defi}

\noindent {\it Remarks:}
 i) If $(P_2,\Pi_2)$ is symplectic, this clearly reduces to
 Def.~\ref{leafsymp}.
 
 ii) Unlike leaf-symplectic embeddings, there always exists a
 leaf-to-leaf symplectic embedding of a Poisson manifold $(P,\Pi)$,
 namely the identity map.

 iii) Non-trivial examples of leaf-to-leaf symplectic embeddings are
 given by second class submanifolds of a Poisson manifold as defined
 in part (iii), (b) of Def.~\ref{coisoPoi} (see also the discussion
 following this definition). Other examples are cosymplectic
 submanifolds \cite{We83} and Dirac submanifolds \cite{Xu} of a
 Poisson manifold; this follows from Corollary 2.11 and Theorem 2.3,
 (vi) of \cite{Xu}.

\smallskip

The situation in Def.~\ref{leaftoleaf} corresponds to a (generalized)
Dirac bracket constructed for a family of second class submanifolds
$C$ in a (nonsymplectic) Poisson manifold $(P,\Pi)$ in the following
way: We use the notation of Remark (iii) at the end of
Sec.~\ref{sec:DirBr}. However, for a degenerate Poisson structure
$\Pi$ there is no symplectically orthogonal complement of
$T_xC$. Instead, we use the fact that $C$ is, as a consequence of
Def.~\ref{coisoPoi}, contained in a leaf $L$ of $P$ with symplectic
structure $\Omega_L$ which defines the complement of $T_xC$ in
$T_xL=T_xC\oplus T_xC^{\perp}$. Similarly, we have
$T_x^*L=\Ann_L(T_xC^{\perp})\oplus\Ann_L(T_xC)$. The projection
$\bar\pi_1$ is now defined in two steps: we first project $T_x^*P$ to
$T_x^*P/\Ann_P(T_xL)\cong T_x^*L$, followed by a projection to the
first factor in the decomposition of $T_x^*L$. Since
$\Ann_P(T_xL)=\ker\Pi_x^{\sharp}$ the Poisson bivector factors through
the first projection 
and $\Pi_D=\Pi\circ(\bar\pi_1\otimes\bar\pi_1)$
defines a bivector which generalizes the Dirac bracket.

\section*{Acknowledgements}

We thank A.~Alekseev, M.~Bordemann, P.~Bressler and in particular
A.~Weinstein for interesting discussions and suggestions, and
L.~Dittmann for help with drawing the diagrams. M.~B.\ is grateful for
support from NSF grant PHY00-90091 and the Eberly research funds of
Penn State, and to A.~Wipf and the TPI in Jena for hospitality during
an essential part of the completion of this work.  T.~S.\ thanks the
Erwin Schr\"odinger Institute in Vienna for hospitality during an
inspiring workshop on Poisson geometry.

\end{document}